\documentclass[aps,prd,twocolumn,superscriptaddress,nofootinbib,floatfix]{revtex4-2}
\usepackage{graphicx}
\usepackage{amsmath}
\usepackage{amssymb}
\usepackage{bm}
\usepackage[colorlinks=true,linkcolor=blue,citecolor=blue,urlcolor=blue]{hyperref}

\begin{document}

\title{A microwave SQUID multiplexing concept for macro-cryogenic calorimeter arrays}

\author{N.~Ferreiro Iachellini}
\email{nahuel.ferreiroiachellini@unimib.it}
\affiliation{Dipartimento di Fisica, Universit\`a di Milano-Bicocca, Milano 20126, Italy}
\affiliation{INFN, Sezione di Milano-Bicocca, Milano 20126, Italy}

\author{P.~Szypryt}
\email{paul.szypryt@nist.gov}
\affiliation{Quantum Sensors Division, National Institute of Standards and Technology, Boulder, Colorado 80305, USA}
\affiliation{Department of Physics, University of Colorado, Boulder, 80309, Colorado, USA}

\author{L.~Canonica}
\affiliation{Dipartimento di Fisica, Universit\`a di Milano-Bicocca, Milano 20126, Italy}
\affiliation{INFN, Sezione di Milano-Bicocca, Milano 20126, Italy}

\author{W.~B.~Doriese}
\affiliation{Quantum Sensors Division, National Institute of Standards and Technology, Boulder, Colorado 80305, USA}

\author{M.~Durkin}
\affiliation{Quantum Sensors Division, National Institute of Standards and Technology, Boulder, Colorado 80305, USA}
\affiliation{Department of Physics, University of Colorado, Boulder, 80309, Colorado, USA}

\author{A.~Giachero}
\affiliation{Dipartimento di Fisica, Universit\`a di Milano-Bicocca, Milano 20126, Italy}
\affiliation{INFN, Sezione di Milano-Bicocca, Milano 20126, Italy}
\affiliation{Department of Physics, University of Colorado, Boulder, 80309, Colorado, USA}

\author{J.~A.~B.~Mates}
\affiliation{Quantum Sensors Division, National Institute of Standards and Technology, Boulder, Colorado 80305, USA}

\author{A.~Nucciotti}
\affiliation{Dipartimento di Fisica, Universit\`a di Milano-Bicocca, Milano 20126, Italy}
\affiliation{INFN, Sezione di Milano-Bicocca, Milano 20126, Italy}

\author{L.~Pattavina}
\affiliation{Dipartimento di Fisica, Universit\`a di Milano-Bicocca, Milano 20126, Italy}
\affiliation{INFN, Sezione di Milano-Bicocca, Milano 20126, Italy}

\author{S.~Quitadamo}
\affiliation{Dipartimento di Fisica, Universit\`a di Milano-Bicocca, Milano 20126, Italy}
\affiliation{INFN, Sezione di Milano-Bicocca, Milano 20126, Italy}

\author{C.~Shiu}
\affiliation{Quantum Sensors Division, National Institute of Standards and Technology, Boulder, Colorado 80305, USA}

\author{J.~N.~Ullom}
\affiliation{Quantum Sensors Division, National Institute of Standards and Technology, Boulder, Colorado 80305, USA}
\affiliation{Department of Physics, University of Colorado, Boulder, 80309, Colorado, USA}

\author{M.~R.~Vissers}
\affiliation{Quantum Sensors Division, National Institute of Standards and Technology, Boulder, Colorado 80305, USA}

\date{\today}

\begin{abstract}
Massive cryogenic calorimeters read out by transition-edge sensors (TES) can reach eV-scale baseline resolution, but the single-channel dc-SQUID readout conventionally used in rare-event searches, with one amplifier chain and several wires per detector, limits practical arrays to a few tens of channels. We propose to apply microwave SQUID multiplexing ($\mu$MUX), developed for fast X-ray and neutrino-mass microcalorimeters, to massive BGO (Bi$_4$Ge$_3$O$_{12}$) calorimeters operated at $\simeq 20$~mK; sapphire and TeO$_2$ absorbers are covered by the same framework. Using the established thermal and noise model for massive TES calorimeters, we derive a noise model for the multiplexed readout, including the HEMT, two-level-system and SQUID contributions, and its scaling with the multiplexing factor $N_{\rm mux}$. Since the slow thermal signals require only about a kHz of sampling per channel, the limiting requirement is not bandwidth but an input-coil sensitivity of $( 0.1$ to $1)\,\mu$A$/\Phi_0$, a factor of $10$ to $30$ beyond current $\mu$MUX devices, matched to the detector current noise of ($9$ to $10$)~pA$/\sqrt{\rm Hz}$, which is independent of the absorber mass. With a total flux noise of $\simeq$$1.2\,\mu\Phi_0/\sqrt{\rm Hz}$, the readout degrades the baseline resolution by less than $2\%$ up to $N_{\rm mux}=1000$, negligible compared with the TES-limited resolution. These results are implemented in a single-tower design: 52 BGO crystals of 100~g each and one HEMT amplifier. The tower is the unit of a multi-tower array of several kilograms aimed at CE$\nu$NS and dark matter searches.
\end{abstract}

\maketitle

\section{Introduction}
\label{sec:intro}

Very low-temperature massive calorimeters are a mature technology for rare-event searches such as dark matter scattering and, at the lowest recoil energies, coherent elastic neutrino--nucleus scattering (CE$\nu$NS)~\cite{Angloher2014,Arnaboldi2008}. Because the lattice heat capacity of dielectric crystals scales with the temperature $T$ as $T^3$, absorbers of $10$~g to $100$~g operated at $\simeq 20$~mK can in principle reach eV-scale baseline resolution~\cite{Pyle2015}. BGO (Bi$_4$Ge$_3$O$_{12}$) is an attractive target for both searches. It is a dense ($7.13$~g/cm$^3$), non-hygroscopic scintillator, and its Bi content makes it the heaviest practical coherent target: the CE$\nu$NS rate grows as $N^2$, where $N$ is the neutron number of the target nucleus, and the spin-independent dark matter rate as $A^2$, being $A$ the mass number of the target. Both $A$ and $N$ are maximal for $^{209}$Bi among usable nuclei. Its Ge content adds $^{73}$Ge, an odd-neutron isotope with nuclear spin $9/2$ and $7.8\%$ natural abundance, which extends the physics case to spin-dependent scattering and, more generally, to the wider set of effective-field-theory operators. At the same time, its light O nuclei receive the larger recoils from sub-GeV dark matter, and the scintillation readout provides event-by-event particle identification. Cryogenic operation of BGO has long been demonstrated: the rare $\alpha$ decay of $^{209}$Bi was first detected with a 46~g BGO scintillating bolometer~\cite{deMarcillac2003}. In the searches considered here, that internal decay serves as a calibration and background tag rather than a target. A tower of $52$ BGO calorimeters of $100$~g each provides $5.2$~kg of active mass, and two or more towers scale the array to $10$~kg and beyond.

The readout, rather than the detector, is today the bottleneck for such arrays. TES-based calorimeters are conventionally read out by direct-current superconducting quantum interference device (dc-SQUID), with one amplifier chain and several wires per channel; at millikelvin temperatures the wiring heat load and the channel count limit practical arrays to a few tens of detectors. Multiplexed SQUID readout, developed over two decades for X-ray and Cosmic MIcrowave Background (CMB) instrumentation, removes this bottleneck. Several techniques are available; a comparative review is given in Ref.~\cite{Akamatsu2022}. In time-division multiplexing (TDM)~\cite{Chervenak1999,Doriese2016}, each TES is coupled to its own first-stage dc-SQUID and the SQUIDs of a readout column are switched on one at a time. Typical factors are $N_{mux} = 30$ to $40$ per column, corresponding to roughly $10$~MHz of usable bandwidth per readout chain and ultimately limited by the $N_{mux}^{1/2}$ aliasing of high-frequency amplifier noise into the signal band; kilopixel X-ray instruments are built from many parallel TDM columns~\cite{Doriese2016}. Code-division multiplexing (CDM)~\cite{Irwin2010} is a Walsh-coded variant of TDM that recovers a $\sqrt{N_{mux}}$ noise advantage but has seen little development beyond its demonstrations. In MHz frequency-domain multiplexing (FDM), each TES is ac-biased at a distinct frequency and the summed currents are read by a single SQUID; factors of $\gtrsim$$40$ over $\sim$$5$~MHz have been demonstrated, bounded by the per-pixel and total SQUID bandwidths. Finally, in microwave SQUID multiplexing ($\mu$MUX)~\cite{IrwinLehnert2004,Mates2011,Mates2017}, each TES is coupled to a dissipationless rf-SQUID that loads a high-$Q$ superconducting resonator in the GHz range; all resonators couple to a single feedline read out by one cryogenic HEMT amplifier, and the response is linearized by flux-ramp modulation~\cite{Mates2012}. Because $\mu$MUX moves the multiplexing problem in the ($4$ to $8$)~GHz band where several GHz of bandwidth are available, it supports factors well above 100: 128 X-ray/$\gamma$-ray TESs have been read out simultaneously~\cite{Mates2017,Noroozian2013}, and the HOLMES experiment has developed a 1024-pixel $\mu$MUX readout for fast neutrino-mass microcalorimeters~\cite{Becker2019}.

To date, $\mu$MUX development has been driven by fast detectors with $\mu$s rise times, which require MHz-scale resonator bandwidths and therefore limit the detector count per readout line. Massive calorimeters sit at the opposite extreme: with ms-scale rise times and sensor bandwidths of a few Hz, each channel requires only kHz of sampling bandwidth. As we show below, this makes multiplexing factors of $100$ to $1000$ on a single coaxial pair natural, provided the SQUID input coupling is adapted to the much smaller current noise of macroscopic calorimeters. The low signal bandwidth would of course raise the achievable factor of any multiplexing scheme, TDM included; the case for $\mu$MUX is what the factor costs, in terms of wiring, at the cold stage. A TDM column of $N_{mux}$ slow detectors would still require $N_{mux}$ first-stage dc-SQUIDs, the associated address lines and their wiring heat load at millikelvin temperature, and would pay the aliasing penalty that grows with column length. In $\mu$MUX, by contrast, the multiplexing factor is carried by one coaxial pair, passive lithographed multiplexer chips, and room-temperature tone generation, so the cryogenic complexity per added detector is essentially zero. Of the established schemes, moreover, only TDM could in principle be extended toward factors of order $1000$ for such slow detectors, and $\mu$MUX holds a further advantage beyond the wiring: its readout chain can be upgraded with near-quantum-limited microwave parametric amplifiers: kinetic-inductance traveling-wave amplifiers combine a system noise of $\simeq$$1$ quantum with a dynamic range three orders of magnitude beyond Josephson devices~\cite{Howe2026}, have been operated at $4$~K as replacements for the semiconductor amplifier in sensor-readout chains~\cite{Malnou2022}, and defer the tone-count noise penalty derived in Sec.~\ref{sec:fluxnoise}, a scaling path that none of the low-frequency schemes possesses. The pressure on channel count, moreover, no longer comes from target mass alone. 

The low-energy excess (LEE)~\cite{Baxter:2025odk} that limits every low-threshold cryogenic calorimeter is currently best attacked at the detector level by reading each absorber with two (or more) independent phonon channels: events originating in the sensor films couple almost exclusively to one channel, while true substrate events are shared between them. Detectors carrying two TES on the same absorber have been built specifically to investigate the origin of the excess, which remains unidentified~\cite{CRESSTDoubleTES2024}, and a two-channel athermal-phonon calorimeter has shown that requiring coincidence between channels selects genuine absorber events, rejecting the single-channel population associated with the sensor films~\cite{AnthonyPetersen2024}. Whichever LEE component dominates in a given device, the approach is the same, more readout channels per crystal, so an array of scintillating crystals is $\times3$ or more in channel count, and high multiplexing factors stop being a convenience and become mandatory. Section~\ref{sec:detector} summarizes the established detector model and the numerical inputs needed for the readout design. Section~\ref{sec:mux} derives the noise of the multiplexed readout and its scaling with the multiplexing factor. Section~\ref{sec:instrument} presents the single-tower instrument, Sec.~\ref{sec:performance} the predicted performance, Sec.~\ref{sec:dm} the projected dark matter reach of one and two towers, and Sec.~\ref{sec:nuem} the sensitivity to neutrino electromagnetic properties and light mediators for targets up to $0.8$~m$^3$ of BGO.

\section{Detector assumptions and readout requirements}
\label{sec:detector}

We do not rederive the calorimeter thermal model. Instead, we use the established framework and design rules of Ref.~\cite{Pyle2015} to calculate the detector noise and signal bandwidth expected in the four configurations obtained by combining calorimetric or bolometric operation with strong or suppressed electro-thermal feedback (ETF). We assume the composite architecture demonstrated in Refs.~\cite{Angloher2014,Angloher2016}: a tungsten TES on a thin dielectric carrier is glued to the absorber, while a gold film provides the weak thermal link to the bath.

The quantities carried into the multiplexing analysis have the following meanings. The model-predicted baseline energy resolution, denoted by $\sigma_E$, is the rms uncertainty with which an optimally filtered, zero-energy event would be reconstructed. The detector current noise is the fluctuation of the TES output current per square root bandwidth within the signal band, before adding readout noise. The bias current is the steady TES operating current and sets the scale of the largest possible pulse. The pulse rise time describes the collection of prompt, nonthermal phonons in the tungsten film, whereas the fall time describes the return of the sensor to its operating point as energy flows to the bath; their inverse time scales define the bandwidth that the readout must preserve. The signal-to-noise bandwidth, denoted by $f_{S/N}$, is the lower-frequency region containing most of the energy-estimator information. The phonon-collection efficiency is the fraction of deposited energy reaching the tungsten film before the crystal thermalizes. Finally, the thermal conductance to the bath controls how quickly the collected energy leaves the sensor and thereby selects between calorimetric and bolometric operation.

\subsection{Athermal-phonon collection}
\label{sec:athermal}

Following Ref.~\cite{Proebst1995}, an interaction first produces high-frequency phonons, of which a fraction, denoted by $\epsilon$, is collected in the tungsten film before thermalization. Let $m$ denote the absorber mass. Scaling the glue and film area as $m^{2/3}$ keeps the assumed collection efficiency approximately mass independent at $\epsilon=1/2$. The prompt-phonon collection time, denoted by $\tau_n$, then scales as $m^{1/3}$ and ranges from $0.6$~ms to $2.9$~ms for 10~g to 1~kg BGO absorbers. The corresponding collection bandwidth, $f_{nt}=1/(2\pi\tau_n)$, ranges from approximately 260~Hz to 55~Hz. The remaining energy reaches the TES as a slower thermal component. These assumptions set the physical pulse rise time and the small-unit resolution scaling used below.

\subsection{Operating modes and ETF regimes}
\label{sec:modes}

\begin{figure*}[t]
\centering
\includegraphics[width=0.92\textwidth]{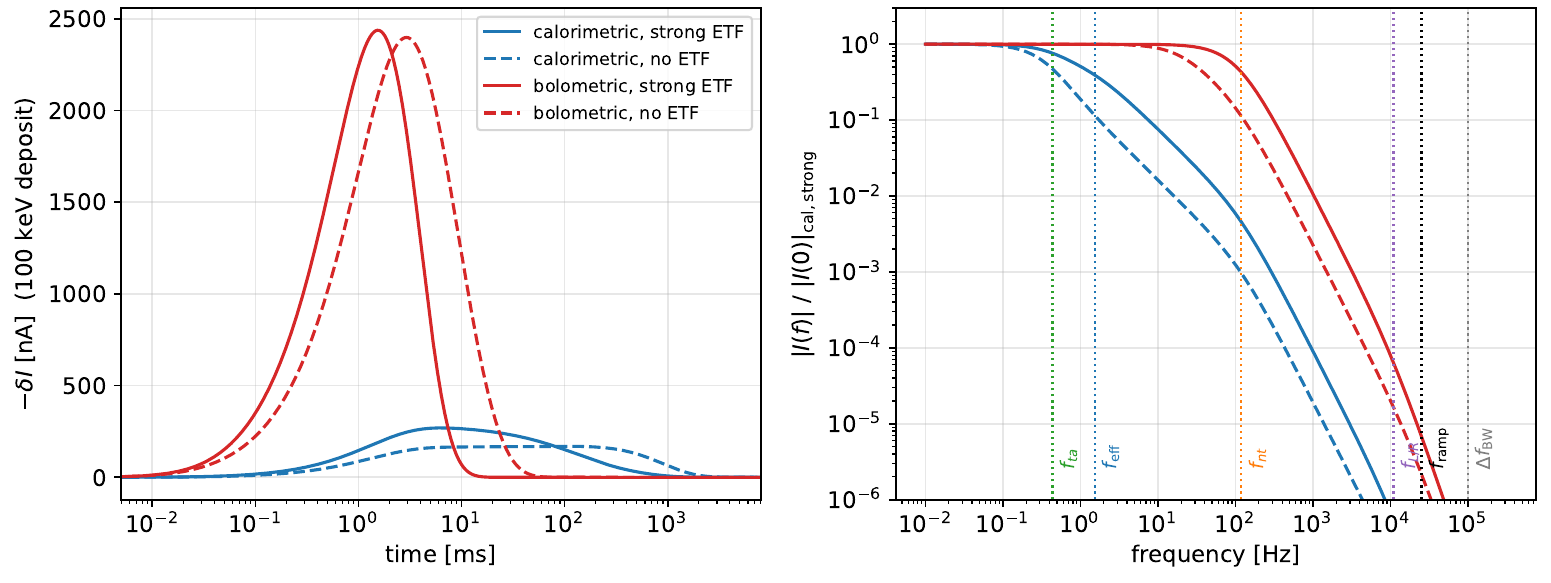}
\caption{Pulse formation for a 100~keV deposit in 100~g of BGO at a TES transition temperature of 15~mK, for calorimetric and bolometric operation with strong or suppressed ETF. Left: predicted current pulses on a logarithmic time axis; the plotted quantity is the decrease in TES current produced by the event. Right: the same pulses in the frequency domain, normalized to their low-frequency response. The vertical markers identify the characteristic frequencies associated with the slow thermal component, the pulse fall, prompt-phonon collection and the electrical response of the TES input circuit. The per-channel sampling rate of 25~kHz and resonator bandwidth of 100~kHz adopted in Sec.~\ref{sec:bw} lie well above the physical signal frequencies.\label{fig:pulses}}
\end{figure*}

\begin{figure*}[t]
\centering
\includegraphics[width=0.88\textwidth]{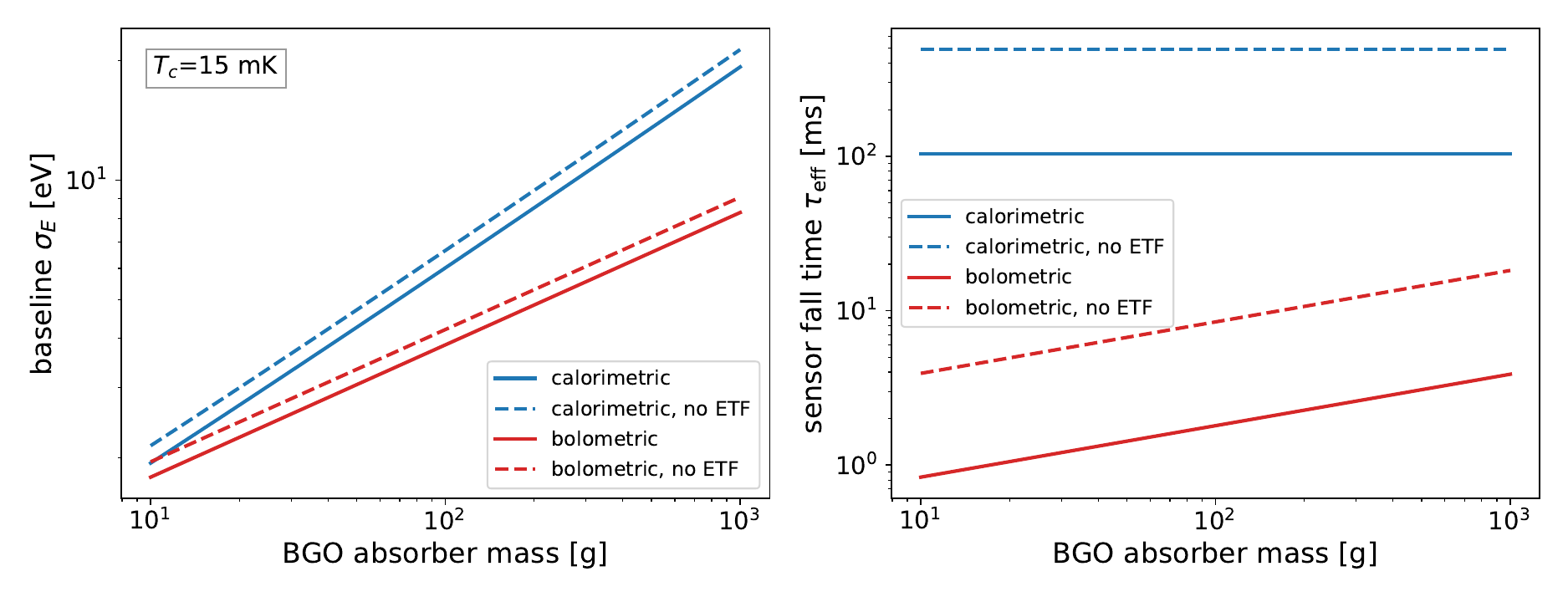}
\caption{Operating-mode comparison at a TES transition temperature of 15~mK in the composite (glued-carrier) architecture for BGO. Left: predicted rms baseline energy resolution versus absorber mass. The bolometric estimate uses the prompt energy collected in the tungsten film and corrects for the assumed 50\% collection efficiency; it overtakes the calorimetric estimate beyond a few tens of grams because it is governed mainly by the sensor-film rather than crystal heat capacity. Right: predicted sensor fall time. Bolometric operation is faster by well over an order of magnitude, and within either mode strong ETF (solid) is about 4.7 times faster than suppressed ETF (dashed), with nearly the same ideal energy resolution.\label{fig:modes}}
\end{figure*}

The two operating modes are selected through the thermal conductance to the bath~\cite{Proebst1995}. In the calorimetric mode, the TES and absorber equilibrate and the deposited energy thermalizes throughout their total heat capacity. This is the simpler and generally more robust choice for small absorbers, whose crystal heat capacity remains low. In the bolometric (athermal) mode, the sensor is made fast enough to measure the phonons collected directly in the tungsten film before the absorber equilibrates. Its resolution is then governed mainly by the much smaller sensor-film heat capacity, denoted by $C_W$, rather than by the crystal heat capacity, making it advantageous for large absorbers, at the price of greater sensitivity to phonon-collection nonuniformity. Strong ETF speeds the response, stabilizes the TES operating point and suppresses current excursions; suppressed ETF gives nearly the same ideal energy resolution because of the conserved gain--bandwidth product, but produces slower pulses and about three times larger current noise. Figures~\ref{fig:pulses} and~\ref{fig:modes} show the calculated pulse shapes, bandwidths and resolution trends for these four configurations. At 100~g and 15~mK the ideal rms resolutions are 6.6~eV (calorimetric) and 4.3~eV (bolometric), with strong-ETF fall times of about 100~ms and 2~ms, respectively. All cases remain far slower than the sampling rate adopted in Sec.~\ref{sec:bw}.

Here strong ETF means the usual voltage-bias feedback: a temperature-induced increase in TES resistance reduces Joule heating and drives the sensor back toward its operating point. In the suppressed-ETF (approximately no-ETF) configuration this self-regulation is intentionally made weak.

\subsection{Absorber materials and design points}
\label{sec:materials}

The absorber heat capacity is evaluated from the low-temperature acoustic Debye model for BGO, Al$_2$O$_3$ and TeO$_2$; the TeO$_2$ calculation agrees with the measured value of Ref.~\cite{Barucci2001}. BGO is the reference target. Sapphire has a much smaller volumetric heat capacity and TeO$_2$ a larger one, but the same readout analysis applies to all three. For BGO absorbers of 10~g to 1~kg at TES transition temperatures, denoted by $T_c$, of (15 to 20)~mK, the model gives ideal rms baseline resolutions of (2 to 43)~eV and a signal-band detector current noise of ($8.9$ to $10.2$)~pA$/\sqrt{\rm Hz}$. The current noise is essentially independent of absorber mass, material and bath conductance and is therefore the principal detector input to the readout design. For the 100~g tower unit, the bias current is (12.7 to 21.9)~$\mu$A over the same temperature range. Existing composite detectors operate about $1.7$ to $10$ above the corresponding ideal resolution~\cite{Abdelhameed2019,Angloher2016}; accordingly, the performance projections use an rms baseline resolution denoted by $\sigma_0=20$~eV and an analysis threshold denoted by $E_{\rm th}=5\sigma_0=100$~eV, within an estimated 40~eV to 200~eV threshold range.

\section{Noise model of the microwave-multiplexed readout}
\label{sec:mux}

In a $\mu$MUX readout~\cite{IrwinLehnert2004,Mates2011}, the TES current, denoted by $I$, produces a flux $\Phi=M_{\rm in}I$ in a dissipationless rf-SQUID, where $M_{\rm in}$ is the input mutual inductance. The change in SQUID inductance with flux shifts the resonance frequency $f_0$ of a dedicated high-quality-factor resonator on the common feedline. A sawtooth flux ramp linearizes the response and is demodulated at the ramp frequency, denoted by $f_{\rm ramp}$, which becomes the per-channel sampling rate~\cite{Mates2012}. The symbol $\Phi_0$ used below denotes one superconducting flux quantum, $S_{21}$ denotes the forward feedline transmission coefficient, and $Q_t$ denotes the loaded resonator quality factor.

\subsection{Flux-noise budget and scaling with the multiplexing factor}
\label{sec:fluxnoise}

Following Refs.~\cite{Becker2019,Mates2011,Akamatsu2022}, the input-referred flux noise of the multiplexer has four components. The HEMT amplifier dominates: for a noise temperature $T_N$ in the range $4$~K to $6$~K and optimal tone power, the theoretical floor including the flux-ramp penalty is $\simeq$$0.7\,\mu\Phi_0/\sqrt{\rm Hz}$, and achieved values are $\sqrt{S_{\Phi,0}^{\rm HEMT}}\simeq(1$ to $2)\,\mu\Phi_0/\sqrt{\rm Hz}$~\cite{Becker2019,Akamatsu2022}. The HEMT also introduces the leading dependence on the multiplexing factor: when the total power of $N_{\rm mux}$ tones approaches the amplifier compression and intermodulation limit, the power per tone must fall as $1/N_{\rm mux}$. Denoting the phase-noise spectrum by $S_\theta$, the power per tone by $P_{\rm feed}$ and the Boltzmann constant by $k_B$, the amplifier contribution scales as $S_\theta \propto k_BT_N/P_{\rm feed}$~\cite{Akamatsu2022},
\begin{equation}
\sqrt{S_\Phi^{\rm HEMT}}(N_{\rm mux}) \simeq \sqrt{S_{\Phi,0}^{\rm HEMT}}\times
\max\!\left(1, \sqrt{N_{\rm mux}/N_c}\right),
\label{eq:hemtN}
\end{equation}
where $N_c$ is the tone count at which the summed power reaches the usable amplifier input and $\sqrt{S_{\Phi,0}^{\rm HEMT}}$ is the per-channel HEMT flux noise value quoted above. 
Naively comparing typical tone ($\sim$$-70$~dBm) and usable input ($\sim$$-45$~dBm) powers gives $N_c\sim300$, but the power that actually loads the amplifier is the transmitted power, reduced by the resonance dips by ($3$ to $6$)~dB depending on the internal quality factor, so the operative knee at optimal readout power is $N_c\simeq600$ to $1200$; we adopt $N_c=900$ in the budget and figures below. Beyond the HEMT, kinetic-inductance traveling-wave parametric amplifiers are the natural upgrade: current devices combine near-quantum-limited system noise ($1.1$ quanta) with a dynamic range three orders of magnitude above Josephson amplifiers~\cite{Howe2026}, and have been demonstrated at $4$~K as alternatives to the semiconductor amplifier in TES-readout chains~\cite{Malnou2022}. A near-quantum-limited noise temperature tolerates proportionally less power per tone at the same flux noise, deferring the compression-driven noise growth to the extent that the amplifier's own saturation power allows, and the kinetic-inductance implementation is the one whose dynamic range moves this trade within reach of large tone combs. We do not assume this upgrade anywhere in the budget below; it is the development direction past $N_{\rm mux}\sim1000$ on a single line, and it has no counterpart in the low-frequency multiplexing schemes. A second tone-count effect with the same scaling deserves mention: the in-band third-order intermodulation products of the tone comb that do not land on resonances form a pseudo-noise floor. Their number grows as $N_{\rm mux}^3/6$, of order $10^{9}$ quasi-randomly distributed in-band products for $2000$ tones, and this floor has proven co-dominant in the multi-thousand-channel $\mu$MUX chains of CMB instrumentation. At the tower scale considered here ($N_{\rm mux}=52$ to $104$ per line) the product count is many orders of magnitude smaller and the mechanism is negligible, but it belongs in the noise budget of any push toward $N_{\rm mux}\sim1000$ on a single line. 

Two-level systems contribute through the resonator frequency jitter. Referred to flux, this noise scales inversely with the frequency swing per flux quantum, which is designed proportional to the resonator bandwidth: narrower resonators are more TLS-sensitive, and the correct evaluation point is the fixed $100$~kHz design bandwidth of Sec.~\ref{sec:bw}. 
There, the conservative bound on the spectral density $S_{\delta f_0}$ of the resonance-frequency fluctuations $S_{\delta f_0}/f_0^2 \lesssim 10^{-19}\,\mathrm{Hz}^{-1}$ at the ramp frequency~\cite{Gao2008,Akamatsu2022}, would allow up to $\sim$$2\,\mu\Phi_0/\sqrt{\rm Hz}$, but multiplexers of exactly this bandwidth class measure total flux noise of ($1.3$ to$2)\,\mu\Phi_0/\sqrt{\rm Hz}$ including the TLS contribution~\cite{McCarrick2021,Groh2024}, showing that the bound is loose at optimal drive power. We therefore adopt a constant allocation of $0.5\,\mu\Phi_0/\sqrt{\rm Hz}$, independent of $N_{\rm mux}$ since the bandwidth is pinned at the fabrication floor. The TLS budget is one more argument for the wider-resonator variant of Sec.~\ref{sec:bw}, whose ($150$ to $200$)~kHz bandwidth reduces this term proportionally. rf-SQUID critical-current and flux $1/f$ fluctuations contribute $\lesssim 0.3\,\mu\Phi_0/\sqrt{\rm Hz}$ at the (high) modulation frequency, and the Johnson noise of the residual input-circuit resistance is negligible~\cite{Becker2019}. Adding in quadrature, the total flux noise is $\sqrt{S_\Phi}\simeq1.2\,\mu\Phi_0/\sqrt{\rm Hz}$, essentially independent of $N_{\rm mux}$ up to $1000$ (Fig.~\ref{fig:ro}, left). The input-referred current noise is then
\begin{equation}
\sqrt{S_I^{\rm ro}} = \frac{\sqrt{S_\Phi(N_{\rm mux})}}{M_{\rm in}} .
\label{eq:siro}
\end{equation}

\subsection{Adapting the SQUID input-coil sensitivity}
\label{sec:coupling}

Equation~\eqref{eq:siro} exposes the central design change required by massive calorimeters. Deployed $\mu$MUX devices use bare rf-SQUID input mutuals of $M_{\rm in}\simeq90$~pH, i.e., an input sensitivity $1/M_{\rm in}\simeq23~\mu$A$/\Phi_0$ and a current noise of ($20$ to $40$)~pA$/\sqrt{\rm Hz}$~\cite{Becker2019}, and current NIST~\cite{10.1063/1.5008527} designs reach $M_{\rm in}\simeq230$~pH ($1/M_{\rm in}\simeq9~\mu$A$/\Phi_0$). This is acceptable for fast X-ray TESs ($\sim$$50$~pA to $100$~pA$/\sqrt{\rm Hz}$) but still above the ($9$ to $10$)~pA$/\sqrt{\rm Hz}$ of the detectors summarized in Sec.~\ref{sec:materials}. We therefore require an input sensitivity
\begin{equation}
0.1~\mu\mathrm{A}/\Phi_0 \lesssim 1/M_{\rm in} \lesssim 1~\mu\mathrm{A}/\Phi_0
\qquad (2~\mathrm{nH} \lesssim M_{\rm in} \lesssim 21~\mathrm{nH}),
\end{equation}
a factor of $10$ to $30$ beyond the current NIST devices. Two implementation routes exist: (i) a much stronger inductive coupler patterned by electron-beam lithography within roughly the same SQUID area, and (ii) a multi-turn input transformer between the TES circuit and the rf-SQUID loop. Both raise the input self-inductance steeply ($L_{\rm in}\propto M_{\rm in}^2$ for the coupler, more for the transformer) and are therefore prone to parasitic microwave resonances. The e-beam-patterned coupler, with the smaller total self-inductance and capacitance, is the natural first step, and the transformer, damped by a lossy filter inserted between the transformer and the SQUID input coil, is the fallback. Sensitivities of this order are routine for dc-SQUID current sensors, whose standard input coils ($L_{\rm in}\simeq350$~nH) correspond to $1/M_{\rm in}$ in between $\simeq0.6~\mu$A$/\Phi_0$ and $1.3~\mu$A$/\Phi_0$ when coupled to typical rf-SQUID parameters. We adopt the upper half of the window, $1/M_{\rm in}$ in $[0.4,1]~\mu$A$/\Phi_0$, as the design point, cap the input inductance at $L_{\rm in}\leq1~\mu$H in the budget below, and defer the implementation to the targeted development of Sec.~\ref{sec:performance}. With $\sqrt{S_\Phi}\simeq1.2\,\mu\Phi_0/\sqrt{\rm Hz}$ this gives $\sqrt{S_I^{\rm ro}}\simeq0.5$~pA$/\sqrt{\rm Hz}$ to $1.2$~pA$/\sqrt{\rm Hz}$, comfortably inside the ($1$ to $10$)~pA$/\sqrt{\rm Hz}$ design envelope (Fig.~\ref{fig:ro}, right).

\begin{figure}[b]
\centering
\includegraphics[width=\columnwidth]{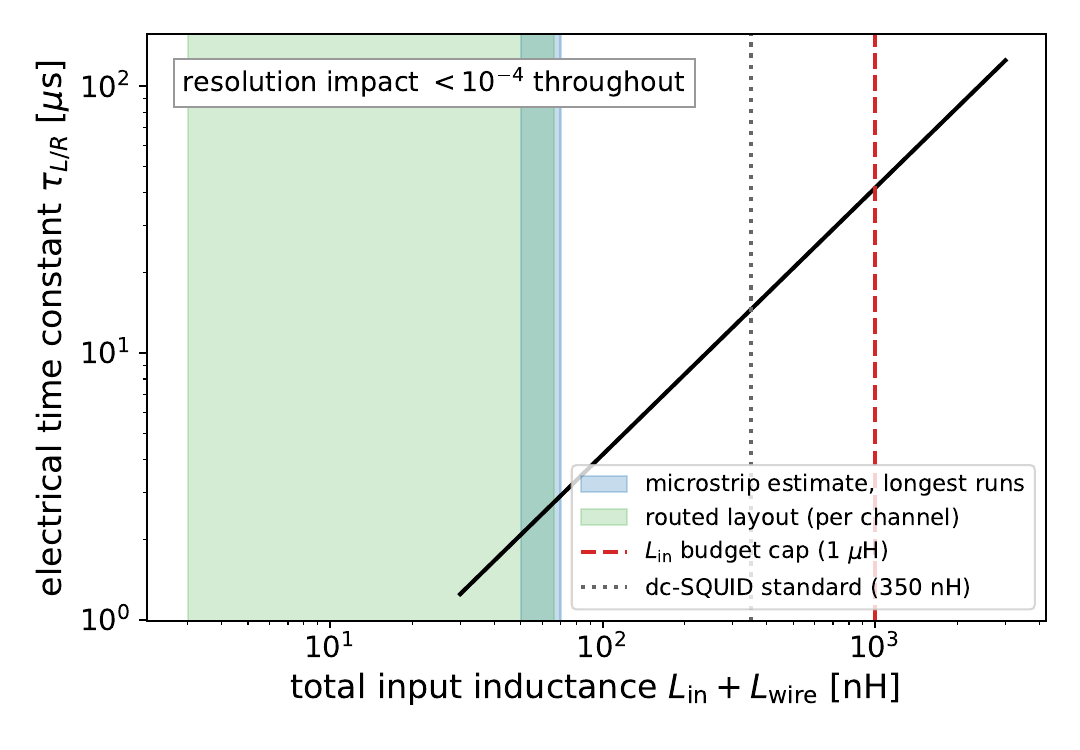}
\caption{Electrical time constant of the TES input circuit versus total inductance $L_{\rm in}+L_{\rm wire}$, with the Nb-wiring contributions (microstrip estimate $\simeq$$50$~nH to $70$~nH for the longest runs, routed layout $3$~nH to $66$~nH) and the $1~\mu$H budget cap marked; the resolution impact is $<10^{-4}$ throughout.\label{fig:ind}}
\end{figure}

The numbers in this paragraph refer to the 200~mm, 52-channel tower geometry of Sec.~\ref{sec:instrument}. Across the wafer, the TES input lines run to the multiplexer chips as superconducting Nb traces a few microns wide. A 4~$\mu$m-wide microstrip placed 1~$\mu$m above the ground plane gives an estimated wiring inductance of 50~nH to 70~nH for the longest (15 to 20)~cm runs; the routed single-layer layout gives 3~nH to 66~nH per channel. We denote the input-coil and wiring inductances by $L_{\rm in}$ and $L_{\rm wire}$, respectively, and cap their sum at 1.1~$\mu$H. At this cap, the TES input circuit has an electrical time constant, denoted by $\tau_{L/R}$, of approximately 46~$\mu$s, corresponding to an electrical-response frequency, denoted by $f_{L/R}$, of approximately 3.5~kHz (Fig.~\ref{fig:ind}). This electrical response filters the signal and detector noise identically, while the readout noise is added afterward at the SQUID. Including it in the optimal-filter calculation changes the baseline resolution by less than $10^{-4}$ over the full inductance range considered because the energy-estimator bandwidth lies far below 3.5~kHz. The thermal response remains about 2000 times slower than the electrical response, preserving electro-thermal stability. The only visible effect is that the electrical edge lengthens from approximately 15~$\mu$s to 46~$\mu$s, which is immaterial for absorber signals but relevant to the direct-sensor-hit veto discussed in Sec.~\ref{sec:etf}.

Two dynamic-range checks follow. First, driving a 100~g TES across its transition changes its current by approximately the 13--22~$\mu$A bias current quoted in Sec.~\ref{sec:materials}. At the proposed input coupling, this corresponds to approximately 15 to 60 superconducting flux quanta at the SQUID input, which flux-ramp demodulation accommodates as a phase winding. Second, a full-scale pulse rising over the 1.3~ms prompt-phonon collection time produces a flux slew of approximately $1.2\times10^4$ flux quanta per second at the preferred input sensitivity of 1~$\mu$A per flux quantum. This remains below the demodulation limit even at a multiplexing factor of 1000. The most aggressive sensitivity considered, 0.1~$\mu$A per flux quantum, would increase the slew tenfold and cap the multiplexing factor near 550, providing an additional reason to adopt the 0.4~$\mu$A to 1~$\mu$A-per-flux-quantum design window.

\subsection{Bandwidth budget}
\label{sec:bw}

The bandwidth budget follows Ref.~\cite{Becker2019} [their Eq.~(B.6)] as a chain of three requirements. Flux-ramp demodulation returns one sample per ramp period, so reconstructing a pulse with rise time $\tau_r$ without resolution loss requires $R_d$ samples on the rise, $f_{ramp}=R_d/\tau_r$; we take $R_d\simeq5$. During each ramp the SQUID sweeps $n_{\Phi_0}=2$ flux quanta and its response is $\Phi_0$-periodic, so the resonance is modulated $n_{\Phi_0}$ times per ramp period and the resonator must track twice that rate, $\Delta f_{\rm BW}=2\,n_{\Phi_0}f_{\rm ramp}$ (with the numbers of Ref.~\cite{Becker2019}: $2\times2\times500$~kHz $=2$~MHz, their resonator bandwidth). Finally, tones are packed at a guard spacing $S=g_f\,\Delta f_{\rm BW}$ with $g_f=7.5$. In practice, narrow-bandwidth resonators require a larger $g_f$: the value $g_f=7.5$ was set to tame the Lorentzian-tail crosstalk of nominally placed MHz-scale resonances, but the fabrication scatter of the resonance placement is fixed in absolute terms, of order a few hundred kHz and largely independent of the designed bandwidth, so resonators in the $100$~kHz to $200$~kHz range still require absolute spacings greater than $1$~MHz to $2$~MHz ($g_f\gtrsim10$). Combining multiplexer chips from different wafer sites (or wafers) adds frequency gaps between chip bands worth another $\sim$$10\%$ on the average spacing. Adding the practical constraint, our only modification, that resonators with bandwidth below $\Delta f_{\rm BW}^{\rm min}\simeq100$~kHz cannot presently be fabricated with adequate yield and $Q$ control, the maximum multiplexing factor over an available readout band $B$ is
\begin{equation}
N_{\rm mux}^{\rm max} = \frac{B}{g_f\,\max\!\left(\Delta f_{\rm BW}^{\rm min},\; 2\,n_{\Phi_0} R_d/\tau_r\right)} .
\label{eq:nmax}
\end{equation}
For fast HOLMES-like detectors ($\tau_r=10~\mu$s) the signal term dominates and Eq.~\eqref{eq:nmax} gives $\sim$$30$ channels per 500~MHz digitizer band. For massive calorimeters with $\tau_r\sim1$~ms the fabrication floor dominates instead: at the nominal $g_f=7.5$, Eq.~\eqref{eq:nmax} would saturate at $N_{\rm mux}^{\rm max}=B/(g_f\,\Delta f_{\rm BW}^{\rm min})\simeq670$ per 0.5~GHz band ($\simeq$$5300$ per 4~GHz HEMT band), while the scatter-driven absolute spacing of $\gtrsim$$2.5$~MHz adopted below reduces this to a practical $\simeq$$200$ per 0.5~GHz digitizer band and $\simeq$$1000$ over the full HEMT band. Even at the floor, each channel offers a sampling rate $f_{\rm ramp}=\Delta f_{\rm BW}^{\rm min}/(2n_{\Phi_0})=25$~kHz, four orders of magnitude above the thermal detector bandwidth. For slow massive detectors, the multiplexing factor is therefore limited not by signal bandwidth but by tone placement and tone count: HEMT compression [Eq.~\eqref{eq:hemtN}] and digitizer dynamic range, both of which enter as the gentle noise growth in Fig.~\ref{fig:ro}.

\begin{figure*}[t]
\centering
\includegraphics[width=0.92\textwidth]{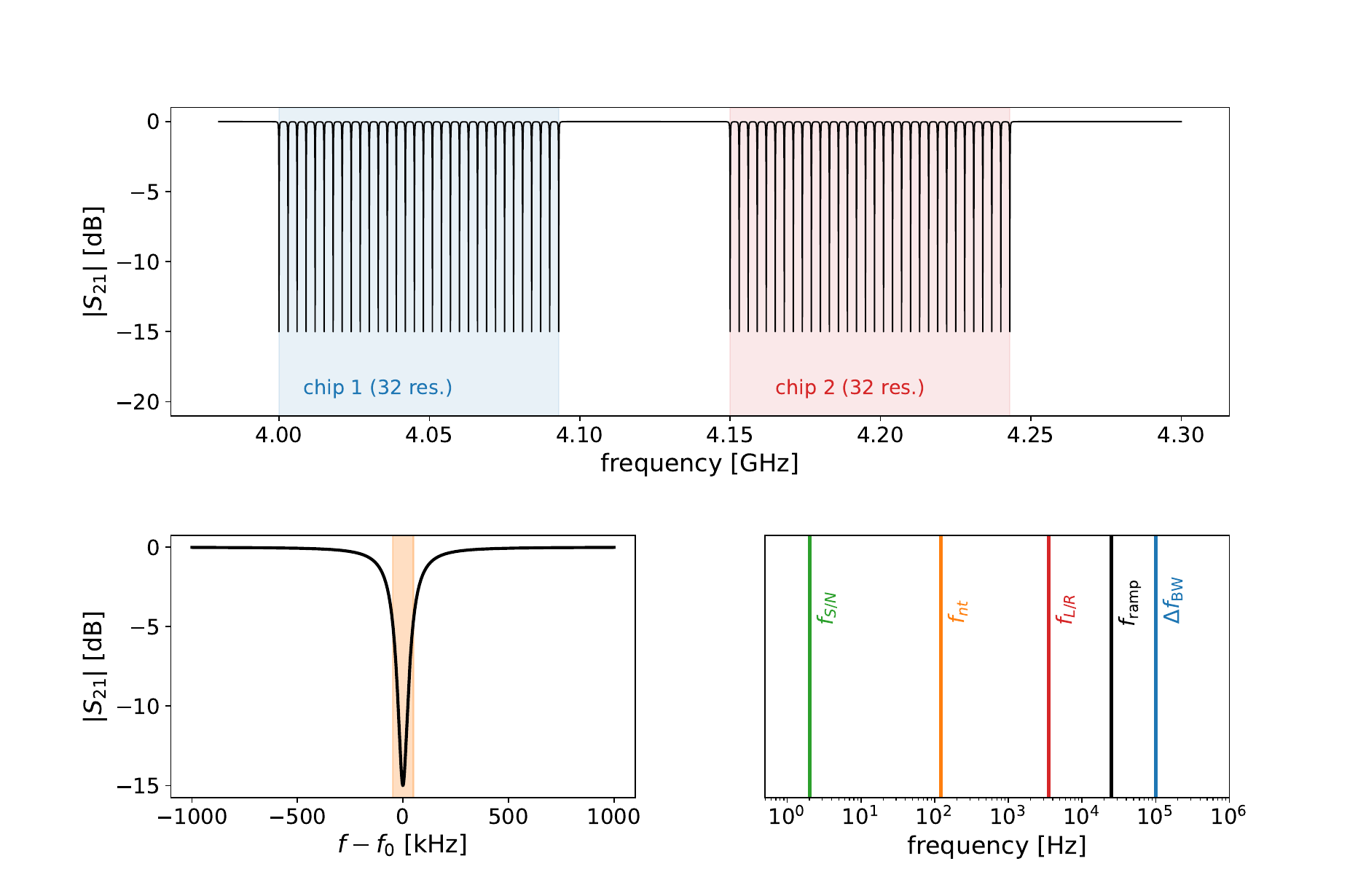}
\caption{Projected readout frequency plan of one tower (cf.\ Fig.~8 of Ref.~\cite{Becker2019}). Top: $|S_{21}|$ comb of the two 32-resonator chips on the single feedline ($3$~MHz spacing, $96$~MHz band per chip, $\Delta f_{\rm BW}=100$~kHz, $15$~dB depth; $26$ channels used and $6$ spares per chip). Bottom left: single resonance at the $100$~kHz floor bandwidth. Bottom right: detector signal bands compared with the per-channel sampling rate $f_{\rm ramp}=25$~kHz and the resonator bandwidth.\label{fig:comb}}
\end{figure*}

Figure~\ref{fig:comb} translates this into the projected frequency plan of one tower, the analogue with our numbers of the measured $S_{21}$ scans of Ref.~\cite{Becker2019} (their Fig.~8): the two 32-resonator multiplexer chips share one feedline at $3$~MHz spacing, four times the $g_f\,\Delta f_{\rm BW}\simeq0.75$~MHz minimum and thus robust against fabrication scatter in the resonance placement, with each resonator at the $100$~kHz floor bandwidth ($Q_t\simeq4\times10^4$). A slightly wider resonator, $150$~kHz to $200$~kHz at the same $2.5$~MHz to $3$~MHz spacing, is an attractive variant: it eases the yield pressure at the fabrication floor, produces deeper resonance dips, and still packs $1000$ to $1200$ channels into the $[4,8]$~GHz band. The surplus per-channel sampling rate ($f_{\rm ramp}=37.5$~kHz to $50$~kHz), unnecessary for the ms-scale rise times here, is better spent sweeping more flux quanta per ramp ($n_{\Phi_0}>2$) and demodulating over more than one $\Phi_0$, which lowers the demodulated readout flux noise: of little value at the tower's tone count, but exactly the free resource that offsets the compression-driven noise growth when pushing toward the tone-count ceiling of a line. The whole tower occupies only $\simeq$$0.2$~GHz, so a full $4$~GHz HEMT band fits $4\,\mathrm{GHz}/3\,\mathrm{MHz}\simeq1300$ resonators (in practice $\sim$$10\%$ fewer once the frequency gaps between chips from different wafer sites are included). This corresponds to $\sim$$40$ multiplexer chips, i.e., $\sim$$20$ towers in principle on one line, and the per-channel sampling rate of $25$~kHz sits two to four orders of magnitude above every detector signal band ($f_{S/N}\simeq2$~Hz, athermal rise $f_{nt}\simeq55$~Hz to $260$~Hz, electrical rise $f_{L/R}\simeq3.5$~kHz). These parameters are close to the $\mu$mux100k devices whose crosstalk was studied in Ref.~\cite{Groh2024} ($1.8$~MHz spacing, $100$~kHz bandwidth): the dominant mechanism there, resonator hybridization between nearest frequency neighbors with amplitude $\chi_f\sim1.5\times10^{-3}$, scales with the bandwidth-to-spacing ratio, so the wider spacing suppresses it further. At this level, crosstalk is negligible for the energy resolution; in a rare-event search it matters mainly as fake coincidences, and is mitigated as in Ref.~\cite{Groh2024} by interleaving, so that frequency neighbors are not physical neighbors, in this case not adjacent crystals.

\subsection{Operating-regime bandwidths}
\label{sec:etf}

The detector model of Sec.~\ref{sec:detector} is used here only to provide the bandwidth and current-noise inputs to the multiplexer. For absorber masses from 10~g to 1~kg, the calculated calorimetric pulse-fall bandwidth is approximately 1.5~Hz with strong ETF and 0.3~Hz with suppressed ETF and is nearly independent of mass. In bolometric operation, the strong-ETF pulse-fall bandwidth decreases from approximately 200~Hz at 10~g to 40~Hz at 1~kg; suppressing ETF makes these pulses about 4.7 times slower. Suppressed ETF also raises the detector current noise to approximately 25~pA$/\sqrt{\rm Hz}$, about three times the strong-ETF value. This relaxes the readout-noise requirement, but the faster and self-stabilizing strong-ETF configuration is preferred for bolometric operation.

\begin{figure*}[t]
\centering
\includegraphics[width=0.92\textwidth]{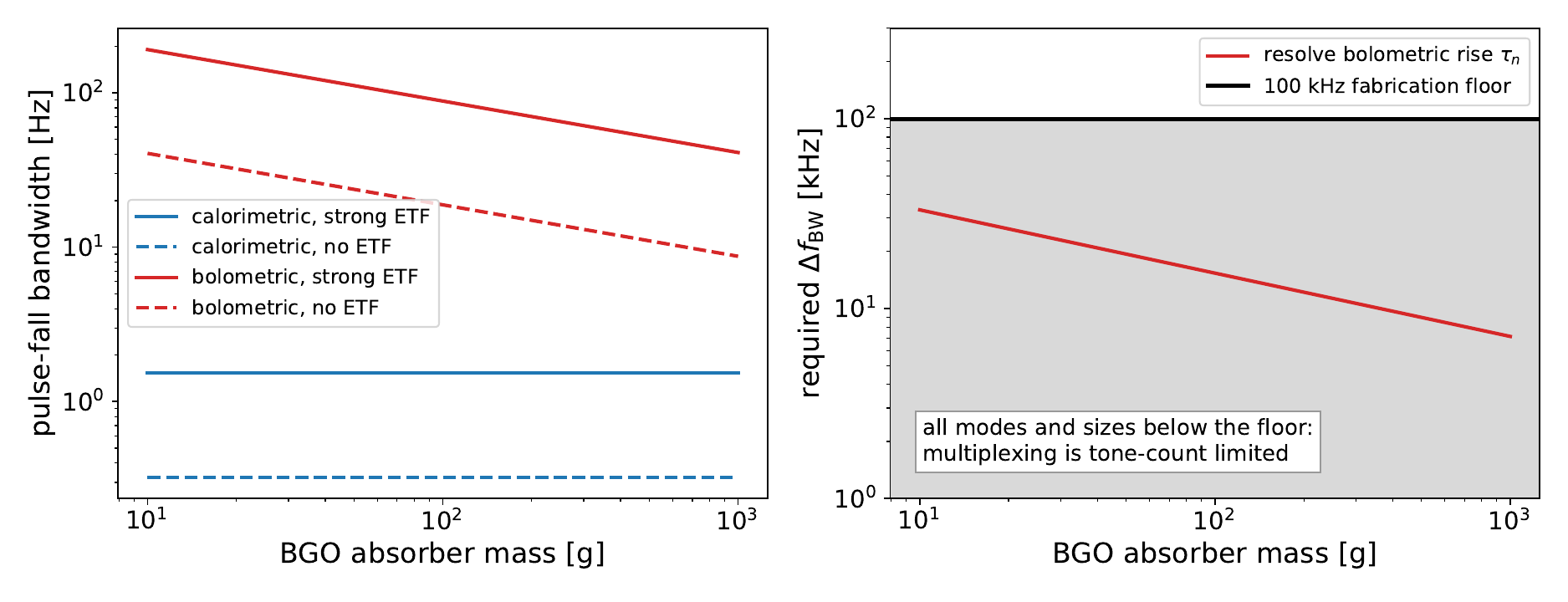}
\caption{Expected signal bandwidth versus absorber mass, operating mode and ETF regime at a TES transition temperature of 15~mK. Left: calculated pulse-fall bandwidth. The calorimetric bandwidth is nearly independent of absorber mass, whereas the bolometric bandwidth decreases as the absorber becomes larger; strong ETF is about 4.7 times faster than suppressed ETF in either mode. Right: minimum resonator bandwidth required to preserve each physical absorber signal, compared with the adopted 100~kHz fabrication floor (shaded). All configurations lie below that floor, so the multiplexing factor is set by resonator placement and tone count rather than detector bandwidth.\label{fig:bwmodes}}
\end{figure*}

\begin{figure*}[t]
\centering
\includegraphics[width=0.88\textwidth]{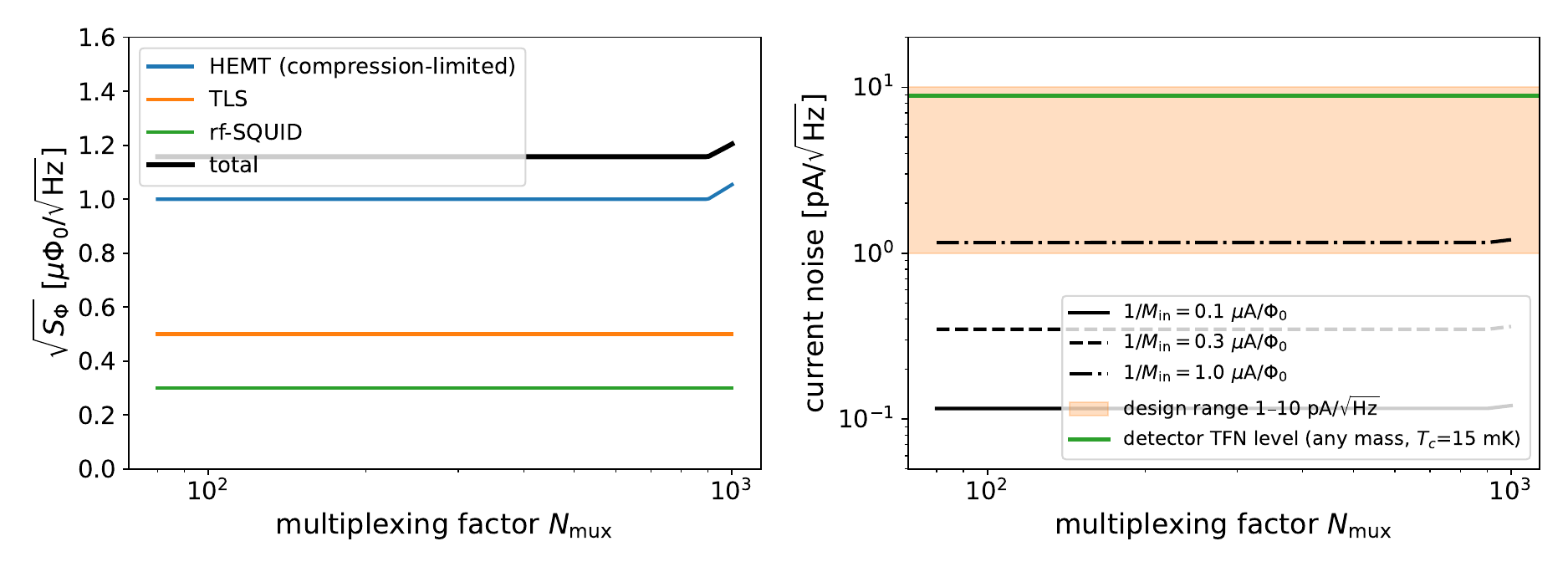}
\caption{Left: flux-noise budget of the multiplexed readout versus multiplexing factor (4~GHz band, $g_f=7.5$, dip-corrected $N_c=900$). Right: input-referred current noise for three input-coil sensitivities, compared with the ($1$ to $10$)~pA$/\sqrt{\rm Hz}$ design envelope (shaded) and the mass-independent detector Thermal Fluctuation Noise (TFN) level (green).\label{fig:ro}}
\end{figure*}

Figure~\ref{fig:bwmodes} translates these detector speeds into the resonator bandwidth that the readout must provide. Across all four configurations, preserving the fastest physical absorber signals requires only about 32~kHz at 10~g and 7~kHz at 1~kg, always below the adopted 100~kHz fabrication floor. The 25~kHz per-channel sampling rate also resolves the 0.6~ms to 2.9~ms prompt-phonon rise with many samples. Consequently, every absorber size and operating regime considered here is limited by resonator placement and tone count rather than detector bandwidth. The only faster feature is the electrical edge of an event occurring directly in the sensor film or carrier. Reconstructing that edge would require approximately 0.44~MHz to 1.3~MHz of resonator bandwidth, depending on the input inductance, and would reduce the multiplexing factor to a few hundred. Such events are a veto class, however, and can be tagged because they rise within one sample; their detailed shape need not be reconstructed.

\section{A single-tower instrument}
\label{sec:instrument}

\begin{figure*}[t]
\centering
\includegraphics[width=0.5\textwidth]{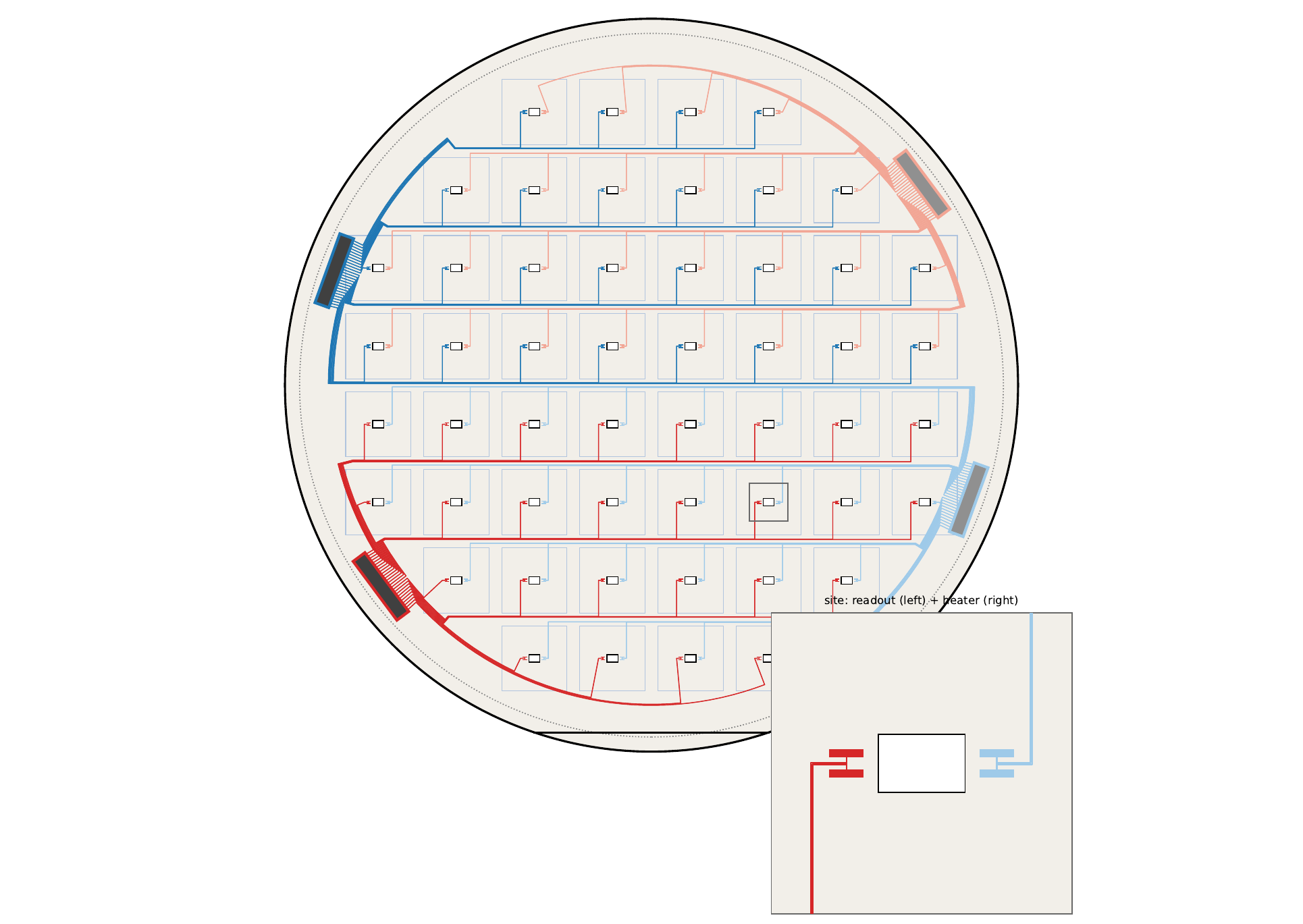}\hfill
\includegraphics[width=0.5\textwidth]{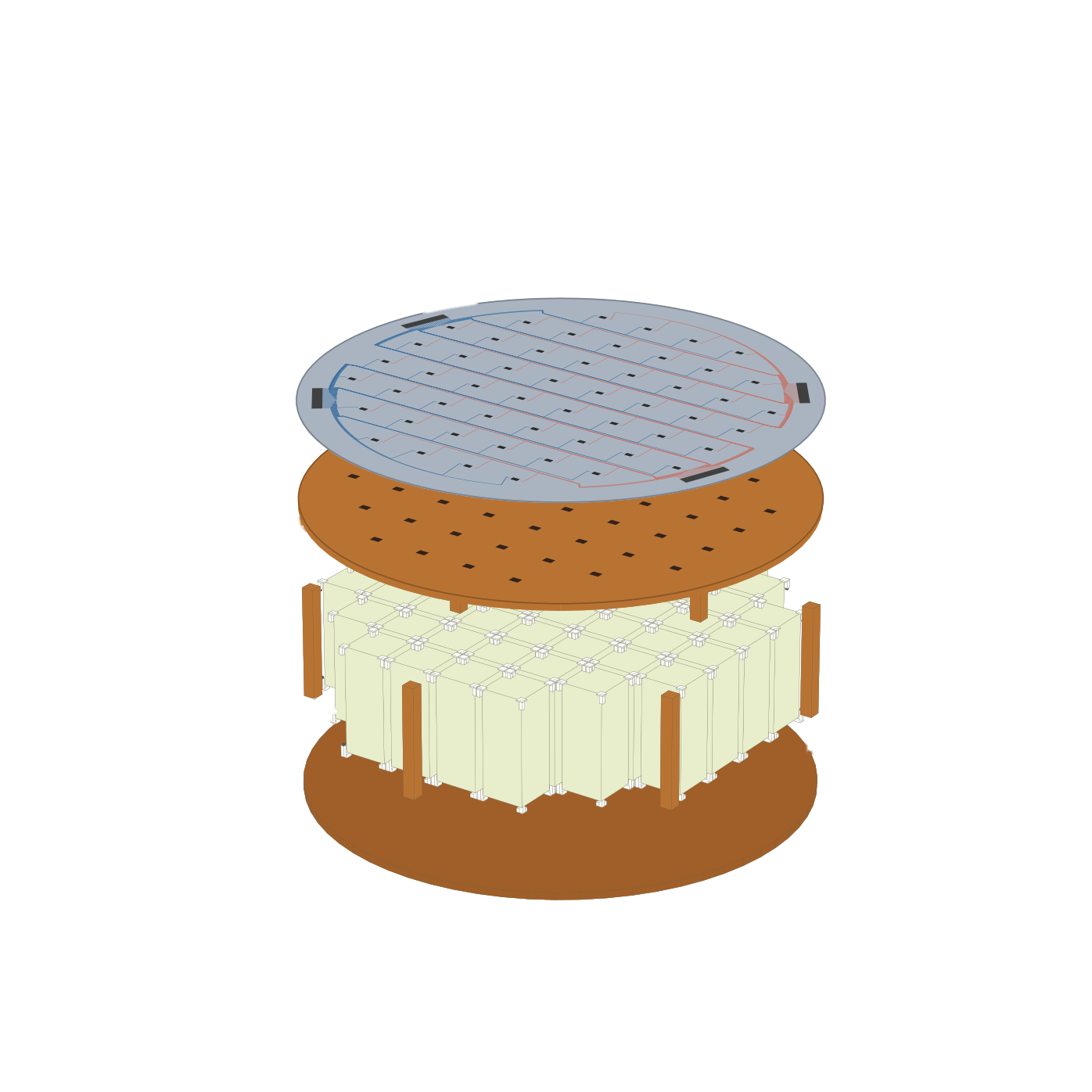}
\caption{The single-tower instrument. Left: single-layer Nb routing on the 200~mm wafer, with 52 sites, the readout network (west chips, dark) and the heater network (east chips, light), and no crossings; the readout bus runs below each crystal row and the heater bus above; inset: one site with the $3\times2$~mm wirebond slot, readout pads (left) and heater pads (right). Right: exploded view, showing the Si wafer, the $3$~mm Cu collar with wirebond feedthroughs, the $52$ BGO crystals in eight PTFE corner L-clamps each, and the $3$~mm Cu base on six standoffs; the stack height is $\simeq$$53$~mm.\label{fig:instrument}}
\end{figure*}

The reference instrument is a single tower of $52$ BGO crystals of $17.8\times17.8\times44.3$~mm$^3$ (aspect ratio $h/a=2.5$) and $100$~g each, for $5.20$~kg of active mass, arranged on a square grid of $21.3$~mm pitch ($3.5$~mm inter-crystal gaps for the clamps) below a $200$~mm (8'') Si readout wafer, $0.5$~mm thick (Fig.~\ref{fig:instrument}). Each crystal carries the glued-carrier composite TES of Sec.~\ref{sec:detector} plus a heater for thermal-gain stabilization, both wirebonded to the wafer through a $3\times2$~mm slot above the crystal, with the TES readout pads to the left of each slot and the heater pads to the right. The elongated crystal shape leaves the athermal design of Sec.~\ref{sec:athermal} essentially unchanged: its volume-to-surface ratio ($V/S=3.7$~mm) is only $8\%$ below that of an equal-mass cube, and the optimal glue area, $\simeq$$60$~mm$^2$ for a 100~g crystal, fits comfortably on the $17.8\times17.8$~mm$^2$ top face.

All wiring is Nb on a single top layer. Two 32-channel $\mu$MUX chips at the west rim of the wafer read $26+26$ channels; two matching chips at the east rim distribute the heater lines. A row-bus architecture routes all $208$ traces (Fig.~\ref{fig:instrument}, left), with a routed per-channel wiring inductance of $3$~nH to $66$~nH, inside the $\leq100$~nH budget of Sec.~\ref{sec:coupling} (resolution impact $<10^{-4}$). The $6$ spare channels of each multiplexer chip remain available for light detectors and dark (witness) SQUIDs. A single coaxial pair and one HEMT serve the tower: $N_{\rm mux}=52$, or $52+52$ when each crystal is paired with a scintillation-light detector in the same band. Either case is far below the HEMT compression knee ($N_c\simeq900$ after correcting for the resonance dips, Sec.~\ref{sec:fluxnoise}) and below the tone-placement ceiling ($\simeq$$200$ per $0.5$~GHz digitizer band at the ($2.5$ to $3$)~MHz pitch, equivalently $\simeq$$1000$ to $1300$ per $4$~GHz HEMT band, Sec.~\ref{sec:bw}), so the readout operates in the regime where its resolution degradation is negligible (Sec.~\ref{sec:performance}). The extra 52 channels for the light detectors could be served with a copy of the Si readout wafer placed at the bottom of the sustaining copper plate. Larger multiplexer chips would simplify the packaging further: 64-channel chips are standard fabrication today, and $\sim$$40$~mm long 128-channel chips appear feasible, so a single chip could in principle serve all $52+52$ heat and light channels of a tower.

The mechanical stack (Fig.~\ref{fig:instrument}, right) consists of, from top to bottom: the Si readout wafer; a solid $3$~mm Cu collar with $4\times3$~mm wirebond feedthroughs above the slots, the crystals sitting fully below it; the crystals, each held by eight PTFE corner L-clamps (four at the top, screwed to the collar, and four mirrored at the bottom, screwed to the base); and a solid $3$~mm Cu base, joined to the collar by six Cu standoffs. The full stack is $\simeq$$53$~mm tall. The tower is the unit of scaling: each additional tower adds $5.2$~kg, one coaxial pair and one HEMT, so a $10$~kg experiment is two towers, and the tone-count ceiling would in fact allow several towers to share one readout line if cabling were at a premium. Chaining towers on one line is not free, however: it requires $\sim$$10$ to $20$ microwave interconnects between towers whose reflections must not build up standing waves on the common feedline. This exceeds current packaging practice, but it is a bounded microwave-engineering problem of impedance-controlled interconnects and feedline equalization rather than a conceptual one.

One cubic meter of instrumented volume is the natural reference point for the readout requirements at experiment scale: it is roughly the largest experimental volume that current cryogenic technology provides, as demonstrated by the CUORE cryostat~\cite{CUOREcryostat}. Allowing $10$~mm of lateral clearance between the $200$~mm towers and $7$~mm between layers for the mechanical frame, about $20$ towers per layer (hexagonally packed) and $16$ layers fit in $1$~m$^3$: roughly $320$ towers and $17\,000$ crystals, i.e., $1.7$~t of BGO at a $23\%$ filling factor. That corresponds to $17\,000$ heat channels, $33\,000$ channels with one light detector per crystal, and $50\,000$ channels if each crystal additionally carries the second phonon channel for LEE discrimination (Sec.~\ref{sec:dm}); at the practical $1000$ to $1300$ tones per $4$~GHz band these are served by roughly $15$, $30$ and $45$ readout lines, i.e., about one HEMT per ten towers. At $1000$ to $1300$ tones per line the intermodulation floor of Sec.~\ref{sec:fluxnoise} enters the budget; it is mitigated by frequency plans that keep the third-order products off the resonances, and in the worst case by retreating to $\sim$$600$ to $900$ tones per line at the cost of $\sim$$30$ additional lines, which changes nothing structural. At these counts the comparison between readout techniques is settled: tens of thousands of channels on a few tens of coaxial lines is a regime no other multiplexing scheme reaches, whatever wiring cost one is willing to pay, and the same architecture extends further by replacing the HEMTs with near-quantum-limited parametric amplifiers~\cite{Howe2026,Malnou2022}, the path toward the high-resolution, low-threshold solid-state detectors of the next generation.

\section{Predicted performance}
\label{sec:performance}

\begin{figure*}[t]
\centering
\includegraphics[width=0.88\textwidth]{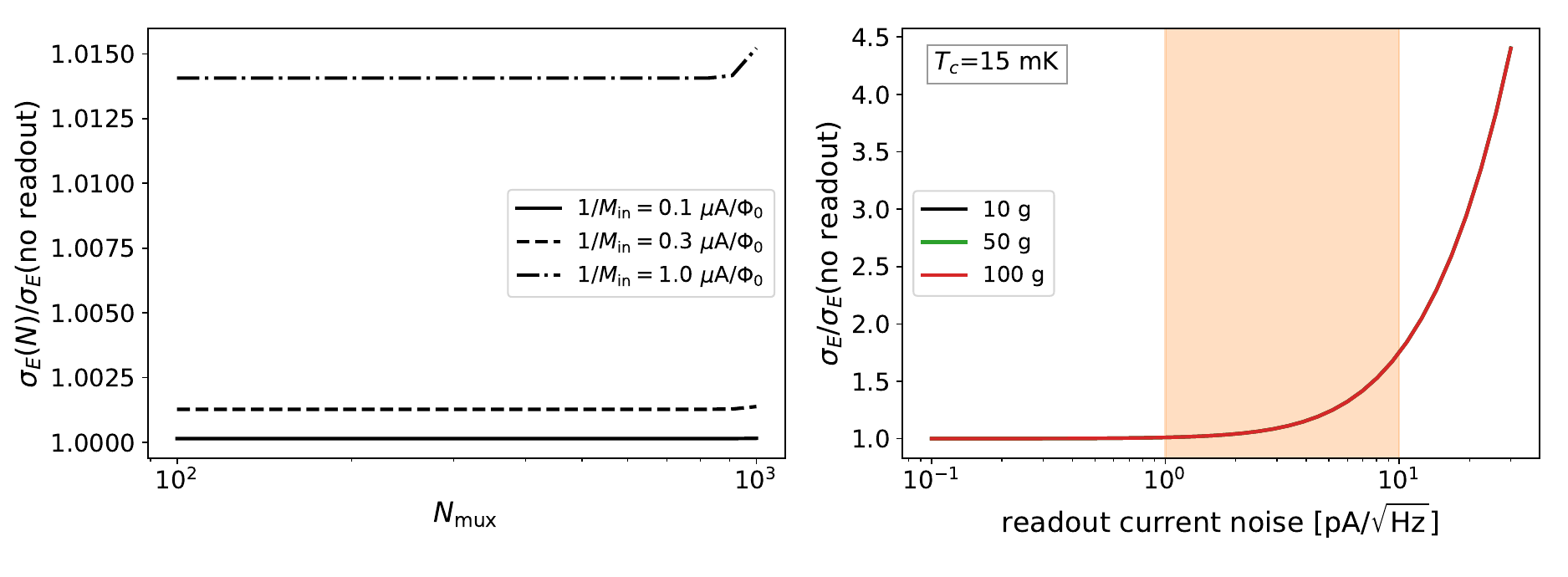}
\caption{Left: baseline-resolution degradation versus multiplexing factor for the model readout noise of Sec.~\ref{sec:fluxnoise} and three input sensitivities (50~g, $T_c=15$~mK). Right: degradation versus a fixed white readout current noise; the curves for 10, 50 and 100~g absorbers overlap exactly, since the detector current noise is mass independent (Sec.~\ref{sec:materials}).\label{fig:deg}}
\end{figure*}

Adding the current noise of Eq.~\eqref{eq:siro} to the detector-noise model, Fig.~\ref{fig:deg} shows the impact of the multiplexed readout on the baseline resolution. With the adapted input coupling ($1/M_{\rm in}\leq1~\mu$A$/\Phi_0$), the degradation is below $2\%$ for any $N_{\rm mux}\leq1000$: the readout is effectively invisible. Even in the pessimistic corner of the design envelope ($10$~pA$/\sqrt{\rm Hz}$), the ideal resolution degrades by only $\sim$$50\%$ (Fig.~\ref{fig:deg}, right), which leaves the $\sigma_E=20$~eV target untouched, since for a detector realistically operating at $\sigma_E=20$~eV the intrinsic noise referred to current is correspondingly larger and the readout contribution relatively smaller.

A single coaxial pair, one HEMT and one digitizer chain can therefore read out $100$ to $1000$ BGO calorimeters, i.e., $10$~kg to $100$~kg of active mass for 100~g units, at TES-limited baseline resolution: a few eV ideal, and $\simeq$$7$~eV to $43$~eV for 100~g units within the demonstrated execution range. The $\sigma_E\simeq20$~eV target requires execution a factor of $\simeq$$5$ above the TES limit of the 100~g unit, inside the range $1.7$ to $10$ demonstrated by operating cryogenic detectors~\cite{Abdelhameed2019,Angloher2016}, and is further relaxed by sapphire absorbers or lighter units (Sec.~\ref{sec:materials}). The tower of Sec.~\ref{sec:instrument} sits well inside this envelope: at $N_{\rm mux}=52$ the flux noise is at its $N$-independent floor and the readout degradation is at the percent level. For scintillating-bolometer operation, the accompanying light detectors (thin Si or Ge wafers with the same TES technology, intrinsically faster and with similar current noise) share the same multiplexer band, using the $6$ spare channels per chip first and then a doubling to $N_{\rm mux}=52+52$ per tower, still an order of magnitude below the tone-count ceiling.

\section{Projected dark matter reach}
\label{sec:dm}

The sensitivity of the instrument to spin-independent (SI) dark-matter--nucleon scattering is evaluated with a binned profile-likelihood analysis: per-bin Poisson counts with a constrained background nuisance per bin, the $\tilde q_\mu$ test statistic with its asymptotic distributions~\cite{Cowan2011}, and median expected 90\%~CL CL$_s$ upper limits from the background-only Asimov dataset.\footnote{The statistical analysis uses the likelihood framework developed for the RES-NOVA dark matter search~\cite{RESNOVA_DM2025}, adapted to the BGO target.} True recoil spectra are taken from \texttt{wimprates}~\cite{wimprates} with the standard halo model and the recommended conventions of Ref.~\cite{Baxter2021} ($\rho_\chi=0.3$~GeV/cm$^3$, $v_0=238$~km/s, $v_{\rm esc}=544$~km/s) and the Helm form factor, summed over Bi, Ge and O weighted by mass fraction; as a pure phonon calorimeter the detector measures the full nuclear-recoil energy, with no quenching. The detector response uses the $\sigma_0=20$~eV execution point defined in Sec.~\ref{sec:materials}: the spectrum is folded with a Gaussian resolution of $\sigma_0$ (truncated at $\pm3\sigma_0$, so that the low-mass reach is not carried by unphysical far tails), cut at the threshold $E_{\rm th}=5\sigma_0=100$~eV, and analyzed in 30 logarithmically spaced bins over $0.1$~keV to $5$~keV.

\begin{figure*}[t]
\centering
\includegraphics[width=0.8\textwidth]{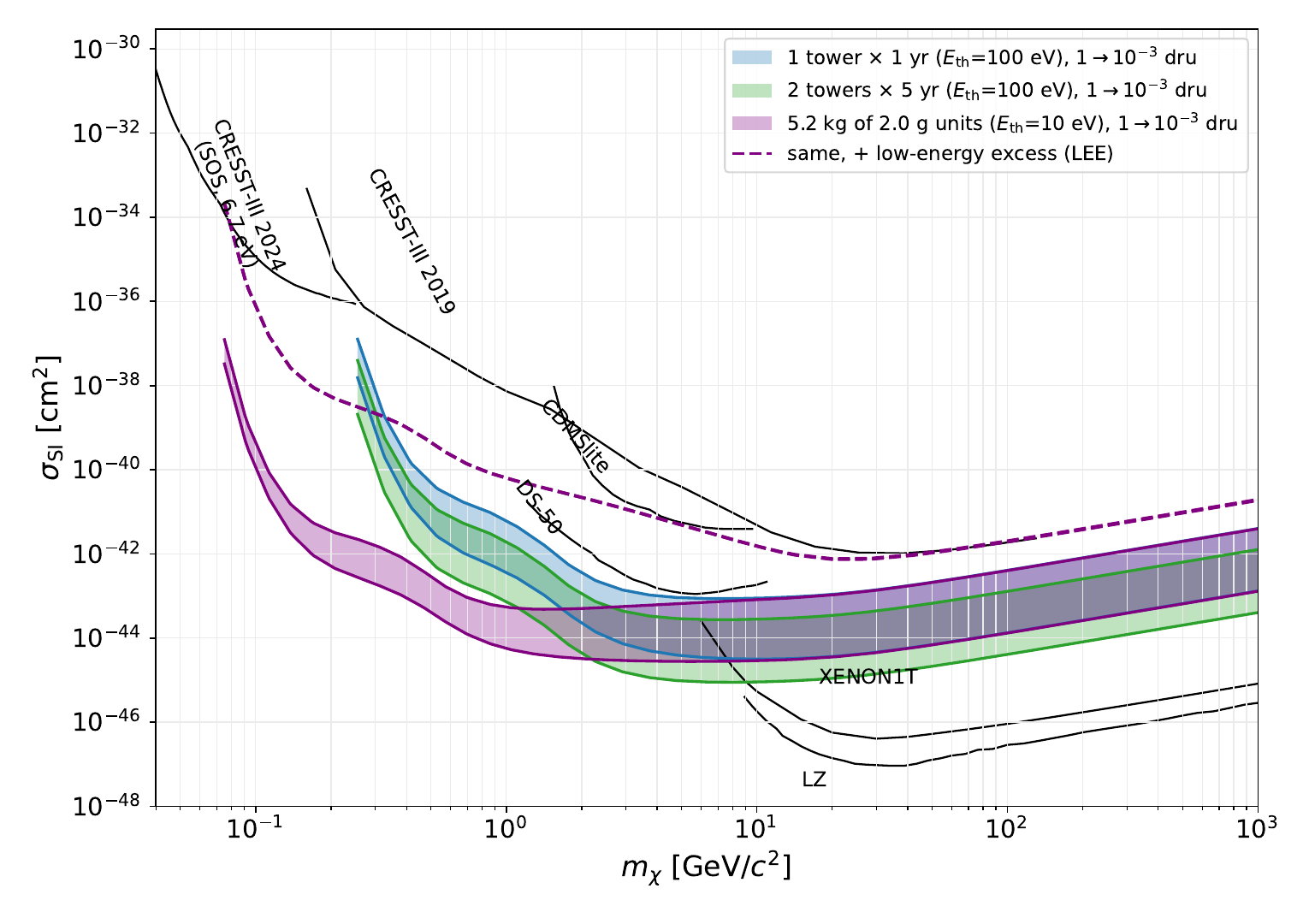}
\caption{Median expected 90\%~CL upper limits on the SI dark-matter--nucleon cross section. Each shaded band spans the background range from $1$~dru (upper edge, no particle identification) to $10^{-3}$~dru after heat--light $e/\gamma$ rejection (lower edge): one tower $\times$ one year ($5.2$~kg$\times$yr, blue) and two towers $\times$ five years ($52$~kg$\times$yr, green), both at $\sigma_0=20$~eV and $E_{\rm th}=100$~eV, and the small-unit array ($5.2$~kg of $2.0$~g BGO units at $\sigma_0=2$~eV, $E_{\rm th}=10$~eV, one year; purple), whose band assumes the flat background only; the dashed purple line adds a low-energy excess of the demonstrated size to its $1$~dru case. The further step to $10^{-4}$~dru gains only a factor $\lesssim$$2$ at these exposures (values quoted in the text) and is omitted for clarity. Black: published limits from CRESST-III 2019~\cite{Abdelhameed2019}, CDMSlite~\cite{CDMSlite2016}, DarkSide-50~\cite{DarkSide2018}, XENON1T~\cite{XENON1T2018}, LZ~\cite{LZ2023} and the $6.7$~eV-threshold silicon-on-sapphire CRESST-III 2024 result~\cite{Angloher2024}.\label{fig:dm}}
\end{figure*}

Two flat background scenarios bound each configuration in Fig.~\ref{fig:dm}: $1$~dru (counts/keV/kg/day), a conservative total rate without particle identification, and a residual $10^{-3}$~dru after the event-by-event heat--light rejection of the $e/\gamma$ population that the scintillating BGO provides (Sec.~\ref{sec:instrument}); each configuration is drawn as the band between the two. A further step to $10^{-4}$~dru would gain only a factor $\lesssim$$2$, since these exposures are statistics limited. The internal $^{209}$Bi $\alpha$ decay deposits $3.14$~MeV, far outside the region of interest, and serves as a tagged calibration line rather than a background. Exposures are one tower for one year ($5.2$~kg$\times$yr) and a two-tower, five-year run ($52$~kg$\times$yr). Figure~\ref{fig:dm} shows the resulting median expected limits. Even the no-rejection scenario improves on the strongest published limits below $\sim$$2$~GeV/c$^2$, where the light O nuclei carry the kinematic reach: $5.0\times10^{-41}$~cm$^2$ at $0.5$~GeV/c$^2$ for one tower-year. Particle identification buys a factor of $\sim$$15$: at $10^{-3}$~dru the tower-year reaches $3.8\times10^{-42}$~cm$^2$ at $0.5$~GeV/c$^2$, $7.0\times10^{-45}$~cm$^2$ at $3$~GeV/c$^2$ and a minimum of $3.2\times10^{-45}$~cm$^2$ near $10$~GeV/c$^2$. The two-tower, five-year exposure at $10^{-3}$~dru reaches $9.0\times10^{-46}$~cm$^2$ at $10$~GeV/c$^2$ and $6.6\times10^{-43}$~cm$^2$ at $0.5$~GeV/c$^2$. Above $\sim$$10$~GeV/c$^2$ the liquid-xenon Time Projection Chambers (TPCs) remain out of reach by orders of magnitude; the natural territory of this instrument is the sub-GeV to few-GeV window, complementary to its CE$\nu$NS program. The comparison with the gram-scale ultra-low-threshold program delimits that window from below: the $0.6$~g silicon-on-sapphire (SOS) detector of Ref.~\cite{Angloher2024}, with a $6.7$~eV threshold, a $1.0$~eV baseline resolution and only $0.138$~kg$\times$d of exposure, holds the strongest limits in the range $74$~MeV/c$^2$ to $202$~MeV/c$^2$~\cite{Angloher2024}, territory kinematically closed to this design, whose $100$~eV threshold cuts the oxygen-recoil acceptance at a dark-matter mass $m_\chi\simeq0.22$~GeV/c$^2$. Conversely, in the overlap region [$0.22$,$0.5$]~GeV/c$^2$ the kg-scale exposure prevails: at $0.3$~GeV/c$^2$ the tower-year improves on the SOS limit by two orders of magnitude already at $1$~dru, and by more than three at $10^{-4}$~dru. Thresholds and exposures, not backgrounds, partition the sub-GeV landscape.

Whether the design would gain by pushing the threshold itself toward that value depends on which scaling law transfers the $0.6$~g demonstration to a $100$~g crystal, and the two available laws describe different design choices in the same (bolometric) mode. The empirical $E_{\rm th}\propto m^{2/3}$ law of Ref.~\cite{Strauss2017} is measured across detectors with a fixed sensor on a growing absorber: the athermal collection efficiency falls with the surface, $\epsilon\propto m^{-2/3}$, and the threshold degrades linearly with the lost amplitude. Under this demonstrated scaling, the SOS threshold does not transfer to large crystals at all, and reaching even $37$~eV would require $\sim$$8$~g units. The optimized-collection architecture of Sec.~\ref{sec:athermal} is designed to evade exactly this. The glue area is co-scaled with the crystal, holding the collected fraction at one half, so the only penalty is the growing tungsten-film heat capacity and $\sigma_E\propto\sqrt{C_W}\propto m^{1/3}$, giving $E_{\rm th}\simeq37$~eV ($\sigma_0\simeq7.4$~eV) at $100$~g, a projection of this design rather than a demonstrated operating point. Note that at equal threshold the mass reach would be the same: in sapphire as in BGO the low-mass acceptance is carried by the oxygen nuclei, and the cutoff scales as $m_\chi^{\rm min}\propto\sqrt{E_{\rm th}}$ ($10$~eV~$\to$~$0.07$~GeV/c$^2$, $37$~eV~$\to$~$0.15$, $100$~eV~$\to$~$0.22$). Extending the acceptance below the tower cutoff therefore requires not a better $100$~g crystal but smaller units. As a concrete design point we take a round $E_{\rm th}=10$~eV, i.e., $\sigma_0=2$~eV at the $E_{\rm th}=5\sigma_0$ rule; denoting the crystal volume by $V$, the optimized-collection scaling ($\sigma_E\propto\sqrt{C_W}\propto V^{1/3}$, since the film scales with the glue area) fixes the unit at $V\simeq0.28$~cm$^3$: $m\simeq2.0$~g of BGO, a $4.8\times4.8\times12.0$~mm$^3$ crystal at the tower aspect ratio. Such units are exactly what the multiplexed readout is for. At the same $3.5$~mm clamp gaps, a $200$~mm wafer carries $\simeq$$350$ of them on a circularly clipped square grid within the same $91$~mm site radius as the tower ($0.71$~kg), read by six 64-channel (or three 128-channel) $\mu$MUX chips on one coaxial line and one HEMT: $\simeq$$350$ tones, inside the practical ceiling of $\simeq$$1000$ to $1300$ per band of Sec.~\ref{sec:bw}, with negligible noise growth below the compression knee. The phonon signals accelerate only mildly ($\tau_n\propto V^{1/3}$, $\simeq$$0.35$~ms), requiring $\simeq$$60$~kHz of resonator bandwidth, still below the $100$~kHz fabrication floor. A tower-equivalent $5.2$~kg is seven to eight such wafers: $\simeq$$2600$ crystals, $\simeq$$45$ multiplexer chips, and three readout lines at $\simeq$$870$ tones each (one line per wafer, $\simeq$$350$ tones, in the conservative variant), a channel count out of reach of any dc-SQUID readout and natural for $\mu$MUX. Figure~\ref{fig:dm} shows this configuration at $5.2$~kg$\times$1~yr on the $1$~dru background, with and without a low-energy excess (LEE) of the size observed in low-threshold cryogenic detectors, fitted from the SOS spectrum~\cite{Angloher2024} as $2.8\times10^{4}\,(E/100~\mathrm{eV})^{-2.2}$~dru, where $E$ is the measured energy. The fit is extrapolated across the ROI. The LEE is irreducible by heat--light identification, since it is a heat-only population, but it is not beyond the reach of the detectors themselves. Reading each crystal with two or more independent phonon channels is the approach the field is converging on: two-sensor detectors have been built specifically to investigate the origin of the excess, still unidentified~\cite{CRESSTDoubleTES2024}, and a two-channel calorimeter has shown that requiring coincidence between channels selects genuine absorber events, rejecting the single-channel population associated with the sensor films~\cite{AnthonyPetersen2024}. Whatever fraction of the LEE ultimately proves discriminable, the cost of finding out is fixed: more channels per crystal. Two phonon channels double the array above to $\simeq$$5200$ readout channels, a regime in which the multiplexed readout is not merely convenient but the only practical option. The kilogram-scale exposure converts directly into reach: even with the LEE included, the array reaches $7.9\times10^{-37}$~cm$^2$ at $0.10$~GeV/c$^2$ and $1.7\times10^{-38}$~cm$^2$ at $0.15$~GeV/c$^2$, extending the acceptance down to $\simeq$$0.07$~GeV/c$^2$; removing the LEE gains a further three orders of magnitude ($4.5\times10^{-40}$~cm$^2$ at $0.10$~GeV/c$^2$). The conclusion mirrors the current experience of the field: below a few hundred eV the collection scaling and the low-energy excess, not the nominal threshold, are the sensitivity frontier, and once they are under control, the multiplexing factor is what turns gram-scale thresholds into kilogram-scale exposures. Two caveats delimit these projections. The sub-GeV curves are exponentially sensitive to the threshold, so the $E_{\rm th}\simeq40$~eV to $200$~eV execution envelope of Sec.~\ref{sec:materials} shifts them accordingly. The readout itself does not enter: at $N_{\rm mux}=52$ to $104$ the multiplexer degrades the baseline resolution at the percent level (Sec.~\ref{sec:performance}), so the reach scales with target mass at constant threshold as towers are added.

\section{Projected sensitivity to neutrino electromagnetic properties and light mediators}
\label{sec:nuem}

The same array is a solar-neutrino detector: elastic $\nu$--$e$ scattering of the pp and $^7$Be fluxes produces electron recoils of energy $E_r$ throughout the region of interest, and any neutrino electromagnetic interaction or new light mediator modifies their low-energy spectrum. The channels below are chosen deliberately. Since the electron-recoil background of this array ($1$~dru) lies orders of magnitude above that of the large xenon TPCs and cannot be reduced by particle identification, the interesting question is where a low threshold on a large target can compensate: the magnetic-moment cross section grows as $1/E_r$ toward threshold and the millicharge as $1/E_r^2$, so these observables measure precisely how far threshold and mass alone carry a background-limited detector. They make the readout argument quantitative: multiplexing is the necessary condition for cryogenic detectors to enter this comparison at all, since it is what turns a low threshold into tonne-scale exposure, but it is not the sufficient one. We evaluate the reach following the XENONnT analysis of Ref.~\cite{Khan2023}: the standard-model $\nu$--$e$ cross section per flavor, the additive magnetic-moment term ($\propto\mu_\nu^2/E_r$), millicharge, charge radius and anapole moment entering through the shift $g_V \to g_V + \tfrac{\sqrt{2}\pi\alpha}{G_F}\,(r_\nu^2/3 - a_\nu/18 - q_\nu/m_e E_r)$, and vector, axial-vector, scalar and pseudoscalar mediators of mass $m_{X}$ with universal couplings, with the flavor decomposition of the oscillation-averaged solar fluxes ($\bar P_{ee}=0.54$, $\theta_{23}$ splitting the $\nu_\mu/\nu_\tau$ conversion). Because the signal itself is an electron recoil, the heat--light particle identification cannot reduce the background here, and all scenarios use the full flat $1$~dru. Bound-electron effects are included in the stepping approximation (electrons with binding energy below $E_r$ contribute, evaluated shell by shell for Bi, Ge and O); the recoil spectrum model reproduces the measured Borexino pp rate~\cite{Borexino2018} to better than $10\%$, and the $20$~eV resolution is negligible on this ROI ($0.1$~keV to $140$~keV, logarithmic binning). Limits are median expected 90\%~CL from an Asimov $\Delta\chi^2$ with a $10\%$ solar-flux pull ($\Delta\chi^2=2.71$ for one parameter; $4.61$ for the two-dimensional coupling--mass contours, as in Ref.~\cite{Khan2023}). Four target scales are considered, from one tower to a RES-NOVA-size volume of $0.8$~m$^3$ of BGO ($5.7$~t): $5.2$~kg$\times$yr, $52$~kg$\times$yr, $0.1$~m$^3\times5$~yr ($3.6$~t$\times$yr) and $0.8$~m$^3\times5$~yr ($28.5$~t$\times$yr).

\begin{figure*}[t]
\centering
\includegraphics[width=\textwidth]{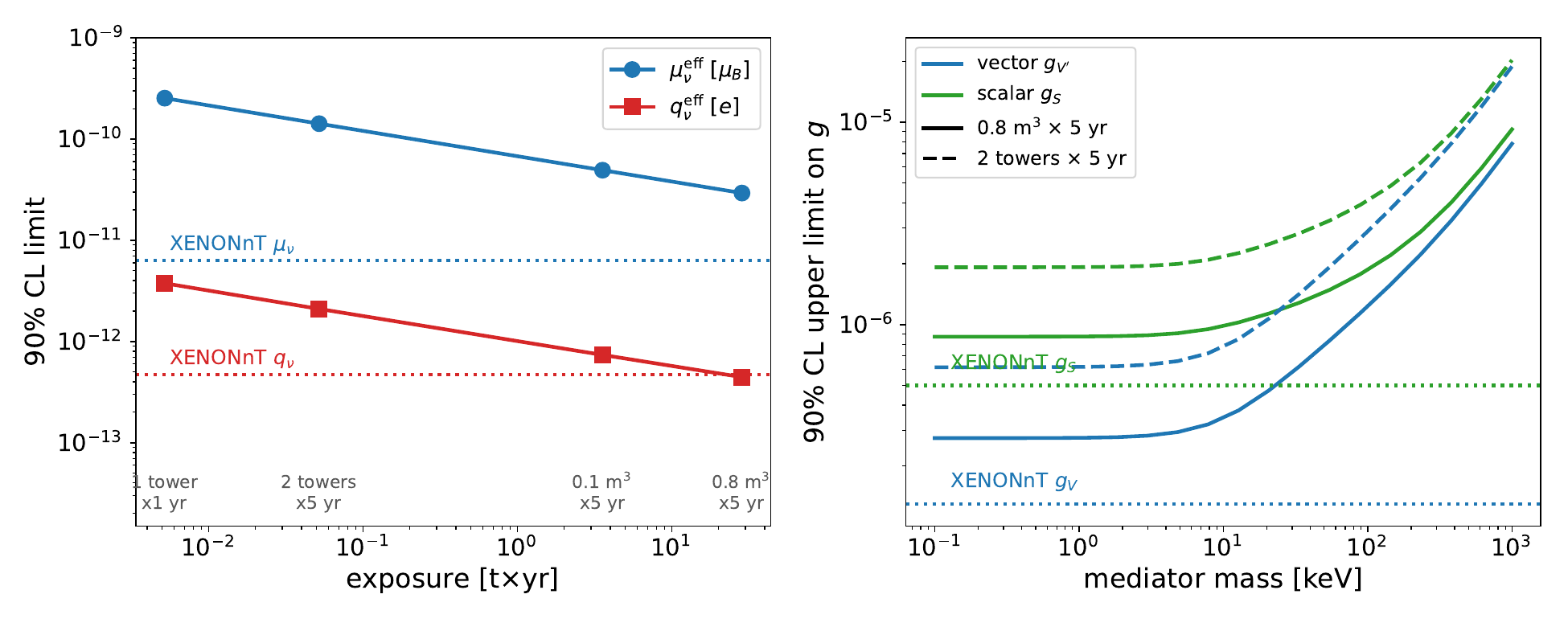}
\caption{Solar-neutrino new-physics reach of BGO arrays with $1$~dru background (analysis of Ref.~\cite{Khan2023} applied to this design). Left: median expected 90\%~CL limits on the effective magnetic moment and millicharge versus exposure, from one tower$\times$1~yr to $0.8$~m$^3\times5$~yr; dotted: XENONnT bounds~\cite{Khan2023}. Right: 90\%~CL upper limits on universal vector and scalar mediator couplings versus mediator mass for the two largest exposures.\label{fig:nuem}}
\end{figure*}

Figure~\ref{fig:nuem} collects the results. Denoting the Bohr magneton with $\mu_B$, the effective (flavor-independent) magnetic-moment limit improves from $2.5\times10^{-10}\,\mu_B$ (one tower-year) to $2.9\times10^{-11}\,\mu_B$ at $28.5$~t$\times$yr, still a factor of $\sim$$5$ above the XENONnT bound of $6.3\times10^{-12}\,\mu_B$~\cite{XENONnT2022,Khan2023}, as expected for a background of $1$~dru against XENONnT's $\sim$$4\times10^{-5}$~dru: in the background-dominated regime the limit scales as $(b/MT)^{1/4}$, where $b$ is the background, $M$ the total mass and T the live-time. It follows that radiopurity is what improves the $\mu_\nu$ reach, more than the mass. The millicharge behaves differently. Its cross section rises as $1/E_r^2$, so the $100$~eV threshold, an order of magnitude below the liquid-xenon ROI, compensates the background handicap, and the $0.8$~m$^3$ exposure reaches $q_\nu \in [-4.1, 4.5]\times10^{-13}\,e$, matching the current world-best solar-neutrino bound $[-1.3,4.7]\times10^{-13}\,e$~\cite{Khan2023}. The charge radius and anapole moment, with no inverse-recoil enhancement, remain one to two orders of magnitude from the xenon bounds ($r_\nu^2 \in [-1.2,0.7]\times10^{-30}$~cm$^2$ and $a_\nu \in [-0.4,0.7]\times10^{-29}$~cm$^2$ at $28.5$~t$\times$yr). Flavor-dependent limits follow the pattern of Ref.~\cite{Khan2023}, a factor of $\sim$$1.4$ ($\nu_e$) to $\sim$$2$ ($\nu_\mu$, $\nu_\tau$) weaker than the effective ones. For the light mediators, the low-mass plateau of the $0.8$~m$^3$ exposure reaches $g_{V'}\lesssim2.8\times10^{-7}$, $g_{A'}\lesssim2.8\times10^{-7}$, $g_S\lesssim0.9\times10^{-6}$ and $g_P\lesssim4\times10^{-6}$ (90\%~CL), within a factor of $\sim$$2$ of the XENONnT bounds; the plateau extends to lower mediator masses than in xenon, since the kernel $1/(2m_eE_r+m_X^2)$ saturates only when $m_X^2 \gtrsim 2m_eE_r^{\rm th}$, i.e., below $\sim$$10$~keV here. The pattern confirms the framing above. Where the cross section rises steeply toward threshold, as the $1/E_r^2$ of the millicharge, the low threshold fully compensates three orders of magnitude of background handicap; where it rises mildly, as the $1/E_r$ of the magnetic moment, it compensates only partially; and where it is flat it does not compete. Multiplexing is thus the technological enabler, admitting a cryogenic array into this comparison by turning its threshold advantage into tonne-scale exposure, but it is not all that is needed: every order of magnitude of background reduction on the electron-recoil floor converts directly into a factor of $\sim$$1.8$ on $\mu_\nu$, and background, not mass, is the remaining lever.

\section{Conclusions}
\label{sec:conclusions}

Microwave SQUID multiplexing, developed for fast X-ray and neutrino-mass microcalorimeters, is well suited to the opposite regime of slow, massive cryogenic calorimeters. We have adapted the thermal and noise model of Ref.~\cite{Pyle2015} to BGO, Al$_2$O$_3$ and TeO$_2$ absorbers of $10$~g to $1$~kg at $T_c$ between $15$~mK and $20$~mK, with heat capacities anchored to the convention-free acoustic floor of the lattice specific heat (validated against the measured TeO$_2$ value~\cite{Barucci2001}) and with the signal collection grounded in the athermal-phonon model of Ref.~\cite{Proebst1995}: a tungsten TES on a glued carrier collects athermal phonons through the epoxy spots, with the glue area scaled to keep the collected fraction at one half and a collection time growing as $m^{1/3}$. We find TES-limited baseline resolutions of $2.0$~eV to $43$~eV (rms) for BGO and a detector current noise of only $9$~pA$/\sqrt{\rm Hz}$ to $10$~pA$/\sqrt{\rm Hz}$, independent of absorber mass ($\sim$$25$~pA$/\sqrt{\rm Hz}$ for the suppressed-ETF bias of Refs.~\cite{Angloher2014,Angloher2016}, which reaches the same resolution with slower pulses and a correspondingly relaxed readout requirement). The derived readout noise model, dominated by the HEMT with a total flux noise of $\simeq$$1.2\,\mu\Phi_0/\sqrt{\rm Hz}$, essentially independent of $N_{\rm mux}$ below the compression knee near $900$, shows that a single readout line can serve $100$ to $1000$ detectors with sub-$2\%$ resolution degradation, provided the SQUID input sensitivity is tightened in the range $[0.1,1]~\mu$A$/\Phi_0$. The bandwidth budget, flux swing and slew rate all close with large margins thanks to the ms-scale signals. The full input-circuit inductance budget (input coil up to $1~\mu$H plus $\lesssim$$100$~nH of on-wafer Nb wiring) affects the resolution at the $<10^{-4}$ level. With the $\simeq$$100$~kHz minimum fabricable resonator bandwidth, energy-only readout is limited by tone placement and count at a practical $N_{\rm mux}\simeq1000$ to $1300$ per 4~GHz HEMT band; even the fastest physical signal, the bolometric collection rise, is resolved at the floor, and the $\mu$s electrical edge of direct sensor hits, a veto class, is tagged rather than reconstructed at the per-channel sampling rate. The composite architecture additionally offers two operating modes on the same hardware: calorimetric, and bolometric (athermal), the latter faster by one to two orders of magnitude and, for crystals beyond a few tens of grams, also the more precise estimator, with strong ETF providing the speed and operating-point stability this mode requires. We have implemented the concept in a single-tower design: $52$ BGO crystals of $100$~g ($5.2$~kg) clamped by PTFE corner clamps between a Cu collar and base below a $200$~mm Si readout wafer, all Nb wiring on a single layer with no crossings ($3$~nH to $66$~nH routed), two 32-channel $\mu$MUX chips and two heater chips on the wafer rim, and one coaxial line and one HEMT per tower ($N_{\rm mux}=52$, or $52+52$ with one light detector per crystal). A profile-likelihood sensitivity analysis of this design (Sec.~\ref{sec:dm}) shows that a single tower-year with $10^{-3}$~dru of residual background after heat--light rejection would probe spin-independent cross sections of $4\times10^{-42}$~cm$^2$ at $0.5$~GeV/c$^2$, beyond current sub-GeV limits even in the $1$~dru no-rejection scenario, with a minimum of $3.2\times10^{-45}$~cm$^2$ near $10$~GeV/c$^2$, and that a two-tower five-year run reaches $9\times10^{-46}$~cm$^2$ at $10$~GeV/c$^2$. As a solar-neutrino detector (Sec.~\ref{sec:nuem}), a $0.8$~m$^3$ BGO array matches the world-best solar-neutrino millicharge bound even at the full $1$~dru electron-recoil background, since the $100$~eV threshold compensates the higher background, while the magnetic-moment reach ($2.9\times10^{-11}\,\mu_B$ at $28.5$~t$\times$yr) scales as $(b/MT)^{1/4}$ and is the physics case for background reduction on the electron-recoil floor. A multi-kg BGO array for CE$\nu$NS and dark matter searches, with a baseline resolution of a few tens of eV (the 20~eV class within the execution range demonstrated by Refs.~\cite{Abdelhameed2019,Angloher2016}), event-by-event heat--light particle identification and two coaxial cables per $5.2$~kg tower is therefore within reach of existing $\mu$MUX technology plus one targeted development: the $10$ to $30\times$ increase in SQUID input sensitivity, via an e-beam-patterned input coupler or a multi-turn input transformer. Near-quantum-limited parametric amplification, with the dynamic range of the kinetic-inductance devices~\cite{Howe2026} and demonstrated at $4$~K as an alternative to the HEMT~\cite{Malnou2022}, but not assumed anywhere in this work, offers a further noise and scaling margin beyond $N_{\rm mux}\simeq1000$ that no low-frequency multiplexing scheme can match. Finally, although every number in this work is evaluated for BGO, nothing in the readout architecture is specific to it: the same wafer, multiplexer and input-coupling design serves the Al$_2$O$_3$ and TeO$_2$ absorbers carried through the detector model, and any other dielectric crystal of interest for direct searches, with the material entering only through its heat capacity and phonon-collection inputs. The channel-count argument will only sharpen: multi-channel phonon readout of each crystal, the demonstrated handle on the low-energy excess~\cite{AnthonyPetersen2024,CRESSTDoubleTES2024}, multiplies the count again, and one cubic meter of instrumented volume at the tower packing corresponds to $17\,000$ crystals and up to $50\,000$ channels on about $45$ readout lines.

\begin{acknowledgments}
This work was supported by the University of Milano-Bicocca through the Bicocca Starting Grant 2026 and by the Istituto Nazionale di Fisica Nucleare (INFN). Doriese, Durkin, Mates, Szypryt, Ullom, and Vissers were supported by funds from NIST.

\textit{This document has not been peer reviewed but has been cleared by NIST for release.}

\end{acknowledgments}

\bibliography{biblio}

\end{document}